\documentclass[12pt]{iopart}

\usepackage{textcomp}
\usepackage{amssymb}
\usepackage{units}

\usepackage[overload]{textcase}

\usepackage{graphicx}
\usepackage{subfigure}

\usepackage{setstack}

\usepackage[sort&compress]{natbib}
\bibpunct{[}{]}{,}{n}{}{}

\begin{document}

\title[Tunable nanopores]{Use of tunable nanopore blockade rates to investigate colloidal dispersions}

\author{G R Willmott$^1$, R Vogel$^2$, S S C Yu$^3$, L G Groenewegen$^3$, G S Roberts$^4$, D Kozak$^4$, W Anderson$^4$, M Trau$^4$}

\address{$^1$ The MacDiarmid Institute of Advanced Materials and Nanotechnology, Industrial Research Limited, PO Box 31310, Lower Hutt 5040, New Zealand}
\address{$^2$ Centre for Biophotonics and Laser Science, School of Mathematics and Physics, University of Queensland, St Lucia, Queensland, Australia 4072}
\address{$^3$ Izon Science Ltd, PO Box 20189, Bishopdale, Christchurch 8543, New Zealand 
}
\address{$^4$ Centre for Biomarker Research and Development, Level 5 East, Australian Institute for Bioengineering and Nanotechnology, University of Queensland, St Lucia, Queensland, Australia 4072}
\ead{\mailto{g.willmott@irl.cri.nz}}

\begin{abstract}

Tunable nanopores in elastomeric membranes have been used to study the dependence of ionic current blockade rate on the concentration and electrophoretic mobility of particles in aqueous suspensions. A range of nanoparticle sizes, materials and surface functionalities has been tested. Using pressure-driven flow through a pore, the blockade rate for 100~nm carboxylated polystyrene particles was found to be linearly proportional to both transmembrane pressure (controlled between 0 and 1.8~kPa) and particle concentration (between 7~x~10$^8$ and 4.5~x~10$^{10}$~mL$^{-1}$). This result can be accurately modelled using Nernst-Planck transport theory. Using only an applied potential across a pore, the blockade rates for carboxylic acid and amine coated 500~nm and 200~nm silica particles were found to correspond to changes in their mobility as a function of the solution pH. Scanning electron microscopy and confocal microscopy have been used to visualise changes in the tunable nanopore geometry in three dimensions as a function of applied mechanical strain. The pores observed were conical in shape, and changes in pore size were consistent with ionic current measurements. A zone of inelastic deformation adjacent to the pore has been identified as critical in the tuning process. 


\end{abstract}



\submitto{\JPCM}
\maketitle

\section{Introduction}

Individual nanopores in thin membranes have been the subject of intense interest over the past ten to fifteen years. They have usually been used to study particles, dispersed in an electrolyte, as they are electrophoretically driven through the pore. A particle within a pore causes a blockade of the ionic current, which can be used to sense or analyse the particle. Much of the recent interest has stemmed from application of this principle to DNA detection, and the prospect of fast sequencing technologies \cite{562,776,611,770,570,572,895}. DNA studies have usually been carried out using the $\alpha$-haemolysin ion channel \cite{589} in a lipid bilayer membrane \cite{611,570,572,770,895}. Solid-state pores composed of silicon \cite{562,776}, track-etched polymers \cite{505,530} or carbon nanotubes \cite{752,580,881,778} have been developed to address the size and stability limitations often encountered with biologically derived nanopores. These synthetic pores can be customised so that their geometry is suitable for sensing bodies ranging from single molecules \cite{562,776} up to macroscopic bodies such as cells \cite{877,767}.


Although a considerable variety of nanopore materials, pore types and length scales have been reported, all of these nanopores have been of fixed geometry. Therefore, each pore is typically useful for detecting a limited size range or type of particle - for example, ssDNA is able to fit through an $\alpha$-haemolysin pore, unlike dsDNA \cite{895}. The response of nanopores to mechanical stimulus (controlled or otherwise) has also not been widely studied, even though the mechanical properties of ion channels are clearly important and potentially useful \cite{884}.



Tunable nanopores (TNs) \cite{454,660,738,771,870,882} have begun to address the topic of mechanically active nanopores. TNs, which are under development by Izon Science (Christchurch, New Zealand), are fabricated in an elastomeric membrane via mechanical puncturing \cite{454}. Membranes can be stretched on macroscopic scales to provide controlled, reversible, nanoscale tuning of the pore geometry, so that a single pore can be used to detect a range of particle sizes. In-situ pore tuning can be used to improve the signal-to-noise ratio for current blockade measurements, to directly study the dependence of signals on pore geometry, or to attempt to clear any obstructions. TNs should enable processes which are not possible using static pores, such as mechanical gating and trapping of particles. 

Previous research on TNs has focused on detection of single dsDNA molecules \cite{454} as well as larger nanoparticles in aqueous suspensions \cite{882,870,738}. The tuning concept has been demonstrated by observing the reversible variation in ionic current at constant applied potential when the membrane is stretched and relaxed \cite{454,660,738}, and the link between macroscopic stretching and nanopore geometry can be quantitatively measured and analysed \cite{660}. The material properties of thermoplastic polyurethane (TPU), the elastomer usually used as a membrane material, have been studied in relation to pore fabrication and tuning \cite{738}. However, a linear elastic model for describing pore stretching \cite{771} has limitations, and there is a lack of data available for developing further mechanical models. Imaging of pore geometry has been limited to two dimensions, so variability from pore to pore and the nanoscale response to macroscopic actuation are not well established.


In this paper, we firstly present a suite of TN images to examine the effects of macroscopic stretching on TN geometry and ionic current. Scanning electron microscopy (SEM) imaging builds on previous work \cite{660,738}, thereby tracking changes in the fabrication and imaging processes, while also identifying ongoing variability between pores. The first three-dimensional (3D) images of a TN, obtained using confocal microscopy, are also presented. Images from both techniques are studied in order to quantitatively link macroscopic applied strain with nanoscale changes in pore size, and the ionic current measured passing through a pore.

Changes in the rate of ionic blockades caused by particles passing through a TN were then used to study (i) particle suspension concentrations and (ii) changes in particle electrophoretic mobility. In the first case, the blockade rate was measured over a range of particle concentrations using pressure-driven flow through a TN. This experiment tests theory predicting a linear relationship between blockade rate and both pressure and concentration, which suggests a method for efficient measurement of unknown particle concentrations. The second experiment uses the TN blockade rate for two sets of silica particles with different size and surface functionality to investigate the effect of particle charge on electrokinetically-driven transport through the TN. The blockade rate is expected to be proportional to particle mobility, and is compared with independent zeta potential measurements. Both experiments employ the blockade rate, presently the most reliable measurement for use with TNs. This research, which concentrates on detection of particles in the range of tens to hundreds of nanometres, bridges the gap between the Coulter approach for cell-sensing and nanopores which detect single molecules. Apart from characterisation of colloids, the particle size range is highly relevant for medicine and public health applications relating to sensing and analysis of viruses such as influenza and norovirus.

\section{\label{PFlux}Theory of Transport through Nanopores}

Measurements of ionic current through a nanopore filled with electrolyte can be compared with a calculated value, using a model in which a potential difference $V_0$ is applied across a membrane, and the transport of ions is entirely electrophoretic. This approach has previously been applied to nanopore work \cite{753,759,769,877,752,778}, including studies of conical nanopores \cite{660,530,503,505}. Assuming that the electric field is uniform across the pore width, that resistance between electrodes is dominated by the pore, and that the electrolyte has uniform resistivity $\rho$, the resistance of a conical pore is

\begin{equation}\label{eq:con}
R_{con}=\frac{\rho l}{\pi ab},
\end{equation}

\noindent where $a$ and $b$ are smaller and larger pore openings respectively (figure~\ref{SetupA}). This expression can be used to calculate geometrical parameters $a$, $b$ and $l$. This theory assumes that no pressure has been applied, and that diffusion is not significant, which is likely to be the case when the electrolyte on either side of the pore is identical.

The experimental blockade rate due to colloidal nanoparticles can be modelled by considering the theoretical transport of particles through nanopores, which is provided by the Nernst-Planck equation. This can expressed by writing the particle flux $\mathbf{J}$ as the sum of terms for diffusion, electrophoresis and convection: 

\begin{eqnarray}\label{eq:NP}
\mathbf{J}&=\mathbf{J_{diff}}+\mathbf{J_{ep}}+\mathbf{J_{conv}} \nonumber\\
&=-D\nabla C+\frac{q}{k_B T}DC\mathbf{E}+C\mathbf{v}.
\end{eqnarray}

\noindent Here, $D$ is the diffusion coefficient, $C$ is particle concentration, $k_B$ is Boltzmann's constant, $T$ is temperature, $\mathbf{E}$ is the electric field and $\mathbf{v}$ is the convective flow velocity. The total effective charge on the particle is $q$. For experimental work, it is most useful to assume that the electric and fluid flow fields are constant, defined using simple geometric models and decoupled from transport mechanisms. This approach 
has been widely applied: to Coulter technology \cite{767}, to particles with length scales of the order of 100~nm \cite{752,778,753}, and to smaller particles \cite{503,530,892}.

For typical nanopore experiments, in which the applied potential is above 0.01~V, (\ref{eq:NP}) suggests that $\mathbf{J_{diff}}< 0.05\mathbf{J_{ep}}$, so that particle diffusion does not significantly contribute to particle transport \cite{870}. When a pressure difference $\Delta P$ is applied across the membrane, and in the absence of other significant body forces, the convective flux is given by the sum of contributions from electro-osmotic and pressure-driven flows. The magnitude of these flows can be determined using the simplified model of a long, thin cylindrical pore of radius $a_0$ and length $l\gg a_0$ \cite{517}. The flux parallel to the cylindrical axis ($J_z=|\mathbf{J}|$) is then the sum of (respectively) electrophoretic, electro-osmotic and pressure-driven terms: 

\begin{equation}\label{eq:terms}
J_{z}=\frac{C}{l}\left(\frac{qD}{k_B T}V_0-\frac{\epsilon \psi_0}{4\pi\eta}A V_0+\frac{a_0^2}{8\eta}\Delta P\right).
\end{equation}

\noindent Here, $\eta$ is the dynamic viscosity of the fluid, $\epsilon$ is the dielectric constant and $\psi_0$ is the potential at the pore wall. The value of $A$ increases asymptotically towards 1, and is greater than 0.9 for $a_0\gtrsim 20$~nm when using 0.1~M KCl, when the double layer thickness is close to 1~nm \cite{564}. It is assumed that electrical resistance between electrodes is dominated by the pore, so that the electric field is $V_0$~/~$l$. (\ref{eq:terms}) is used as the basis for interpreting the experiments in this paper. Note that all three terms scale linearly with particle concentration $C$, while the predicted dependence of particle flux on applied pressure is linear. 

When pressure and a potential are both applied across the membrane, the ratio of pressure-driven particle flux to electrophoretic particle flux is

\begin{equation}\label{eq:rat1}
\frac{J_{z,pressure}}{J_{z,ep}}
=\frac{3 \pi a' a_0^2 \Delta P }{4 q V_0}.
\end{equation}

\noindent Assuming a pore radius of 1~$\mu$m, an applied potential of 0.3~V, a particle of hydrodynamic radius $a'=50$~nm (diameter 100~nm) and estimating the effective charge $q$ using a typical value for carboxylated polystyrene colloids (9.6~x~10$^{-3}e$~nm$^{-2}$ \cite{780}), this ratio is greater than unity when $\Delta P$ exceeds $\sim$120~Pa. For experiments in Section~\ref{presstext}, transmembrane pressures in the range 0 to 1.8~kPa are applied, so pressure is therefore expected to dominate electrophoretic particle transport.

When there is no pressure difference across the pore, only electro-osmotic to electrophoretic fluxes contribute to particle transport. The ratio of these terms is


\begin{equation}\label{eq:rat2}
\frac{J_{z,eo}}{J_{z,ep}}
=-\frac{3 a' \epsilon \psi_0}{2 q}A.
\end{equation}

\noindent For water at 293~K, a pore wall surface potential of -75~mV (a value previously used for track-etched polymers \cite{522,535,671}), and determining $q$ as described above, this ratio is -0.04~A and -0.02~A for particles of hydrodynamic radius $a'=100$~nm and $a'=250$~nm respectively. It is therefore expected that electrophoresis dominates electro-osmosis in Section~\ref{zeta}. In electrophoretic transport, the particle mobility $\mu$ is defined by 

\begin{equation}\label{eq:zeta}
\mu=\frac{qD}{k_B T},
\end{equation}

\noindent and Smoluchowski's approach \cite{891} gives the particle zeta potential as

\begin{equation}\label{eq:zeta2}
\zeta=\frac{\eta\mu}{\epsilon}.
\end{equation}

\section{\label{MM}Materials and Methods}

\subsection{Particle Sensing Apparatus}

\begin{figure}
\begin{center}
\subfigure[]{\label{SetupA}\includegraphics[width=5.5cm]{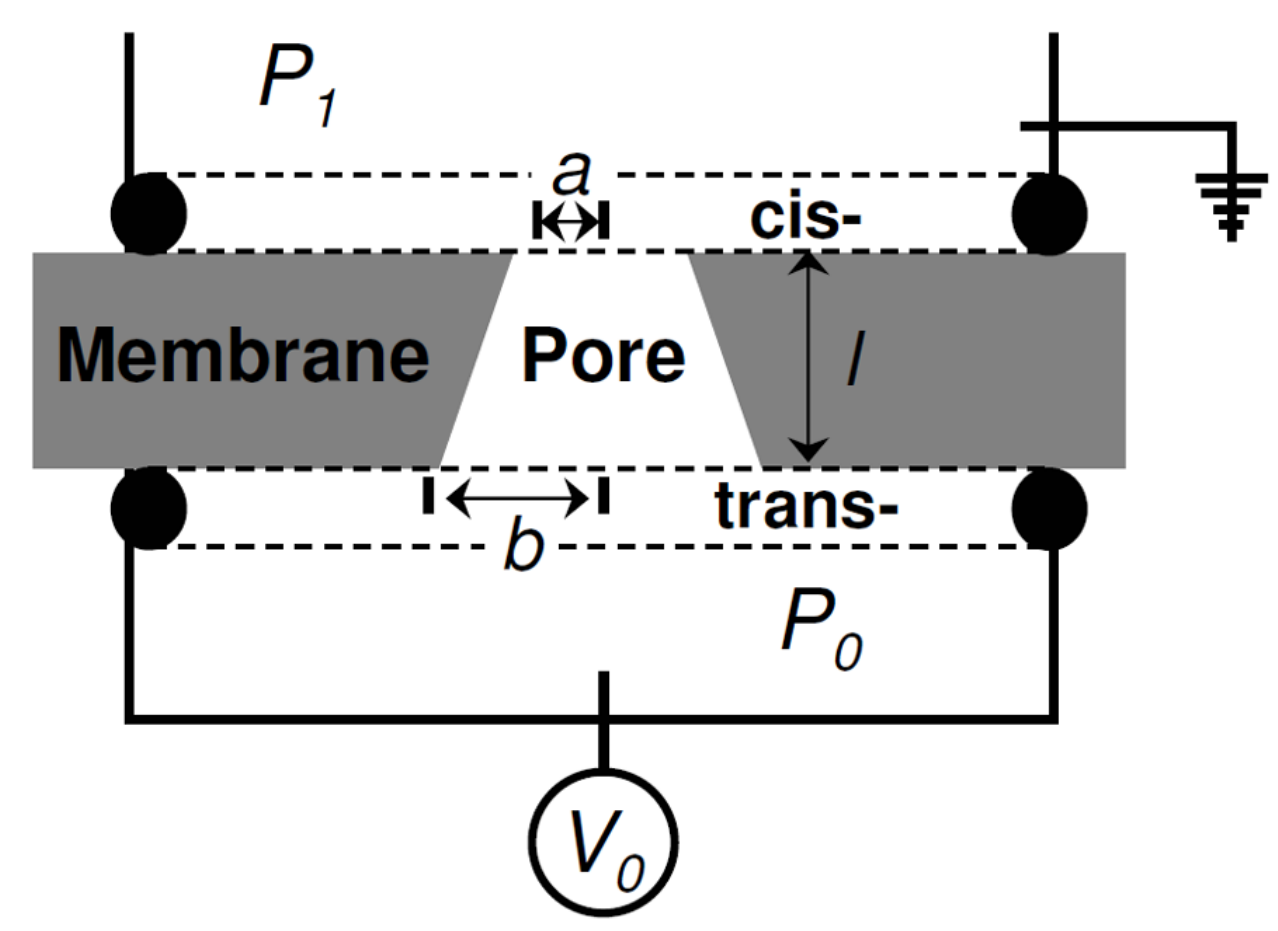}}
\subfigure[]{\label{SetupB}\includegraphics[width=5.5cm]{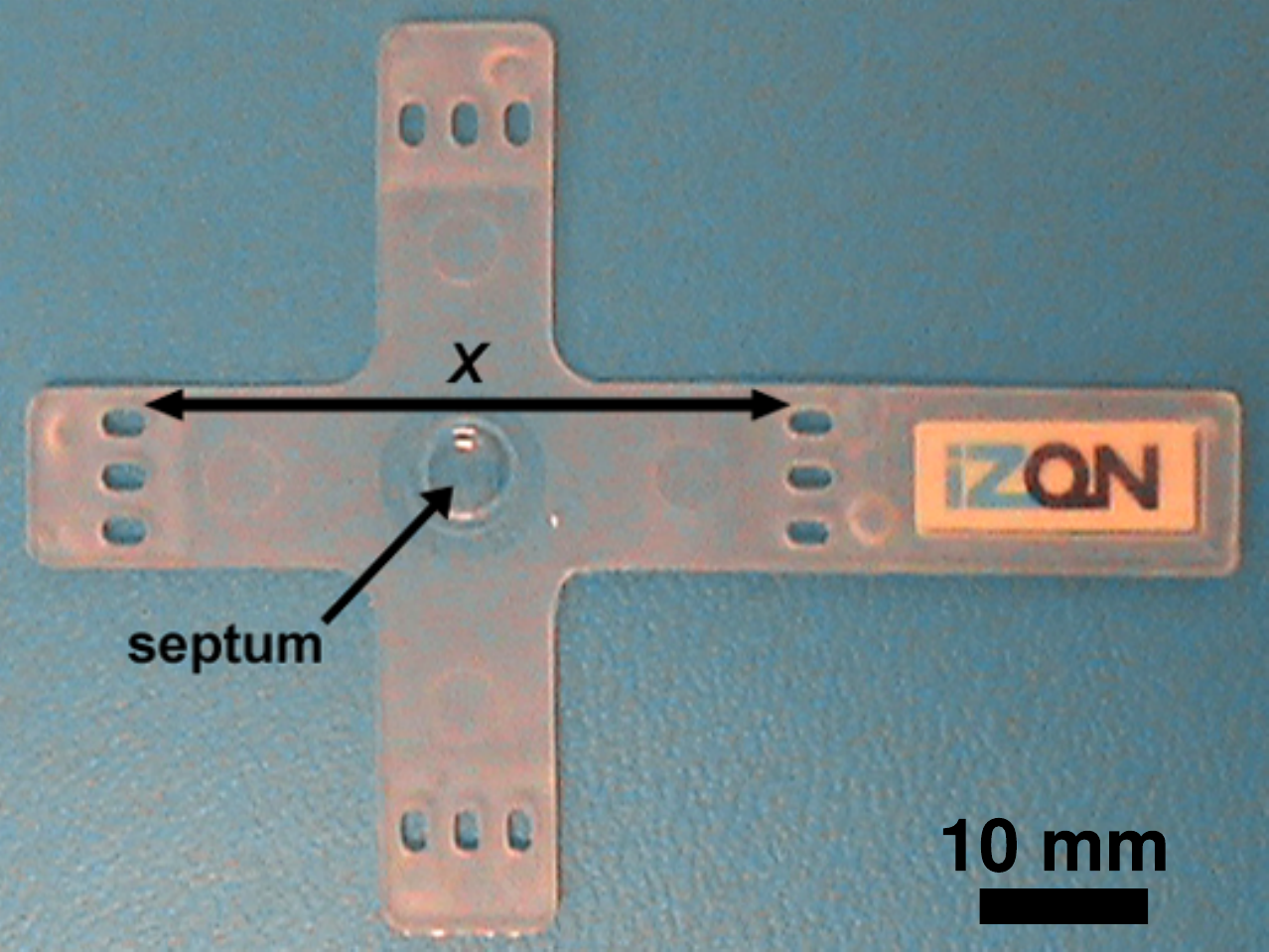}}
\subfigure[]{\label{SetupC}\includegraphics[width=5.5cm]{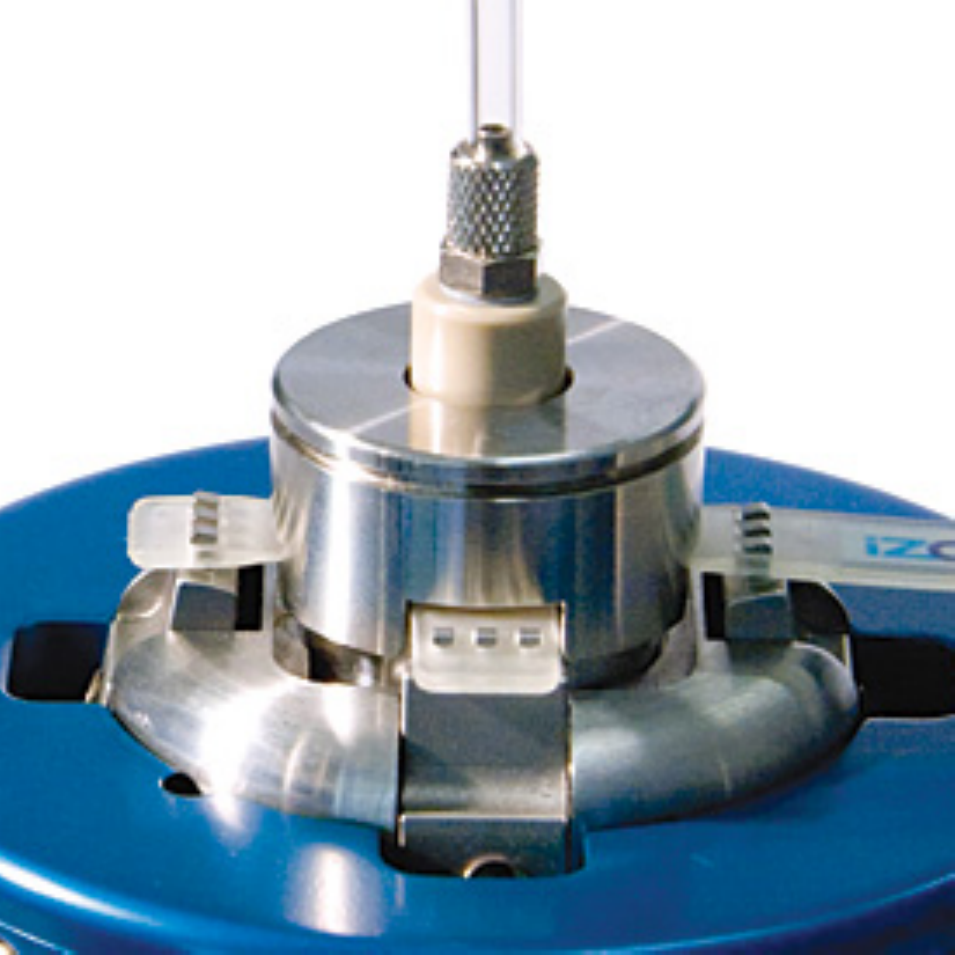}}
\subfigure[]{\label{SetupD}\includegraphics[width=5.5cm]{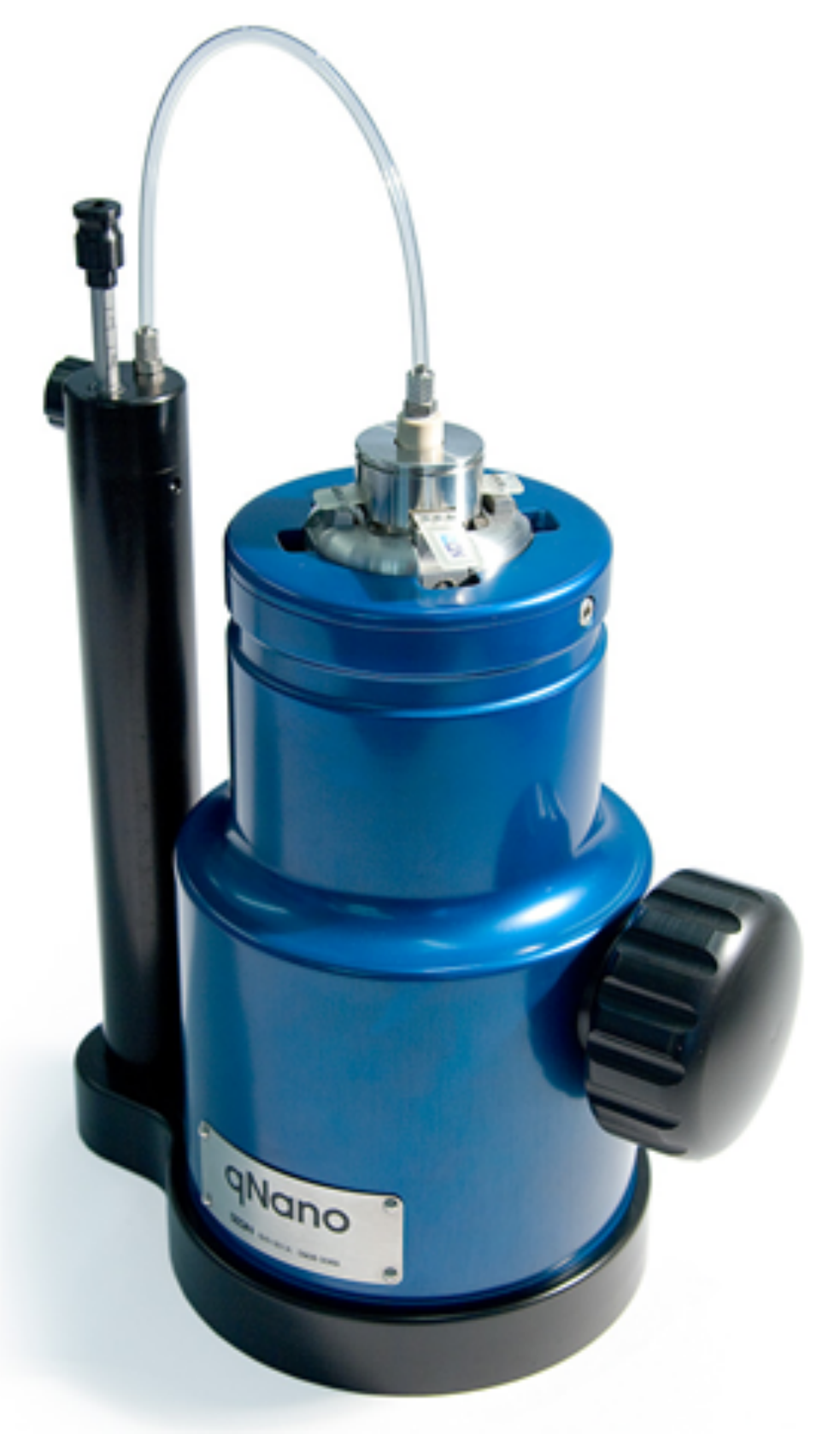}}
\end{center}
\caption{(a) Schematic sectional diagram showing a conical pore between two halves of a fluid cell, symmetric about a cylindrical axis running through the centre of the pore. Labelled parameters are described in the text. (b) A TPU cruciform of thickness 0.7~mm, increasing up to 1.5~mm around the holes at the ends of the cruciform legs. (c) In use, the cruciform is housed within a fluid cell, with the legs protruding and the holes placed on pegs to enable stretch tuning. (d) The fluid cell is part of the Izon qNano apparatus, which includes customised electronics in the base for precise ionic current measurements; a handle is turned to stretch the cruciform; and a manometer column, which can apply pressure to the upper fluid cell via the transparent polymer tubing.}
\label{Setup}       
\end{figure}

TNs are formed by mechanically puncturing a 200~$\mu$m-thick TPU membrane (Elastollan 1160D, BASF) using a chemically-etched tungsten needle \cite{454}. Pores generated are approximately conical in shape (figure~\ref{SetupA}), and taper from the larger puncture hole on the `cis-' side of the membrane, to a smaller hole on the `trans-' surface. The size and shape of the pore can be controlled by altering the tungsten needle etching or puncturing processes. All pores in this paper use similar pore fabrication protocols and are therefore comparable. 

Membranes are supported as a thin septum at the centre of a cross-shaped stretching platform known as a `cruciform' (figure~\ref{SetupB}), which is formed by thermal injection of TPU into a mould. Mechanical stretching is achieved using pegs placed in the holes at the ends of the cruciform legs. For quantitative analysis of macroscopic stretching, we define $X$ as the distance between outer edges of the pegs placed in holes on opposite legs of a cruciform (figure~\ref{SetupB}). Only symmetric stretching, in which $X$ is identical across each pair of legs, has been studied. $X$ also defines an effective macroscopic strain parameter:  

\begin{equation}\label{eq:alpha}
\alpha=\frac{X-X_0}{X_0},
\end{equation}

\noindent where $X_0$ is the value of $X$ with no stretching applied. In the present study, $X_0=40$~mm was consistently used. 

For ionic current measurements, the cruciform is placed in the middle of a fluid cell (figure~\ref{SetupC}) fitted on the Izon qNano apparatus (figure~\ref{SetupD}). Ag/AgCl electrodes in the upper and lower fluid cells are used to apply potential difference $V_0$ across the membrane and measure the ionic current through the pore. The ionic current is measured using electronics built into the qNano device. Data is digitized and interpreted using Izon's customised v.1 instrument control software.  

A U-tube manometer incorporated into this arrangement (figures~\ref{SetupC} and \ref{SetupD}) applies a pressure head $\Delta P = P_1-P_0$ across the membrane, enabling pressure-driven particle transport. Pressure is established by connecting one end of the U-tube to the top of the fluid cell and keeping the other end open. By adjusting the water levels in the tubes, a net pressure is applied to the top fluid cell, driving the particles through the nanopore from top to bottom. 

\subsection{Solutions and Particles}

A standard electrolyte buffer (SEB) of 0.1~M aqueous KCl, 0.01~M tris(hydroxymethyl) aminomethane (Tris, pH 8) and $<$~0.1~w/w\% Triton X-100 was used in all experiments. For experiments in Section~\ref{presstext}, a chelating agent (0.1~w/v\% EDTA) was added. For all experiments, the nanopore and cell were cleaned with SEB, then the cell was loaded with 40~$\mu$L of suspended particles. Suspensions were made by diluting a stock solution of particles in SEB at the desired pH. For the experiments in Section~\ref{presstext}, carboxylated polystyrene particles with nominal average diameter of 100~nm and as-received concentration of 1.8~x~10$^{13}$~mL$^{-1}$ were sourced from Thermo Scientific. These particles were diluted in SEB and ultrasonicated for at least 5~minutes prior to use. Experiments were set up as shown schematically in figure~\ref{SetupA}: the nanopore was oriented with the small opening facing upwards. The solution containing nanoparticles was introduced into the top half of the fluid cell with plain buffer solution in the lower half and a positive (downward) pressure applied ($P_1>P_0$). 

For Section~\ref{zeta}, amine functionalised 200~nm diameter silica particles were purchased from Micromod Partikeltechnologie (Germany). Carboxylic acid functionalised 500~nm diameter organosilica particles were synthesised in a two step condensation process according to the methods outlined by Miller et al. \cite{888} and Vogel et al. \cite{889}. Briefly, 3-mercaptopropyl trimethoxysilane (Lancaster, UK) was hydrolysed overnight in acidified water, followed by overnight condensation with triethyleamine. The particles were then reacted overnight with 3-aminopropyl trimethoxysilane and triethyleamine in ethanol \cite{890}. Surface amine groups were carboxylic acid modified by overnight reaction with succinic anhydride in anhydrous dimethylformamide. All reagents were purchased from Sigma Aldrich unless otherwise stated. 

\subsection{Other Methods}

Scanning electron microscopy (SEM) was used to image surfaces of two TPU membranes containing pores. Cruciforms were coated with 5~nm of gold/palladium prior to imaging with a JEOL 6700 field emission SEM system. Imaging results depend on the behaviour of this coating as well as the characteristics of the particular pore. 

Confocal microscopy was used to construct 3D images of the pore geometry as a function of strain. The cruciform was firstly loaded onto the tuning platform with the trans- surface facing downwards (upside-down with respect to figure~\ref{SetupA}). Standard electrolyte was used in the bottom half of the fluid cell and a fluorescent dye solution (10$^{-5}$~M Rhodamine B (Sigma Aldrich) in SEB) was loaded into the top half. The pore was stretched open to $X = 47.5$~mm, and the dye solution was left to run through the pore for approximately 150~minutes. The membrane was then dried in a nitrogen stream before imaging using a confocal microscope (Zeiss LSM 710, with LD Plan-Neofluar 20x/0.4 Korr M27 objective) with laser excitation of wavelength 514~nm. Z-stack images were compiled using images taken in the $x$-$y$ plane every 3~$\mu$m along the $z$-axis, with the vertical range starting 10~$\mu$m below the first visible fluorescence and ending 10~$\mu$m above the last visible fluorescence. The z-stacks were reconstructed into 3D and cross-sectional images using Zeiss ZEN 2008 software. The technique is non-destructive: ionic current measurements can be made after the imaging process.

For the blockade rate experiments in Section~\ref{presstext}, the applied potential was $V_0=0.3$ and the nanopore was stretched to $X= 42.5$~mm. In Section~\ref{zeta}, nanopore stretch and cell voltage were tuned at pH 8 for each set of particles (amine or carboxylic acid coated silica) to give an approximate blockade rate of 1000 particles per minute. This corresponded to $V_0=0.3$ and 1~V and $X= 59.7$ and 45.5~mm for the 500~nm carboxylic and 200~nm amine silica particles, respectively. These settings were used for all additional sample runs with varying pH, besides switching the polarity of $V_0$ when the particles passed through their isoelectric point. The silica particle size and zeta potentials were measured using a Malvern Zetasizer 3000 HSA in SEB, with pH adjusted between 3 and 10 using 0.1~M~HCl and 0.01~M~NaOH. The pH was measured before and after particle addition to ensure no change had occurred. 

\section{\label{2.2}Imaging and Stretching of Pores}

\subsection{Scanning Electron Microscopy} 

\begin{figure}
\begin{center}
\subfigure[]{\label{SEM1Cis}\includegraphics[width=5.5cm]{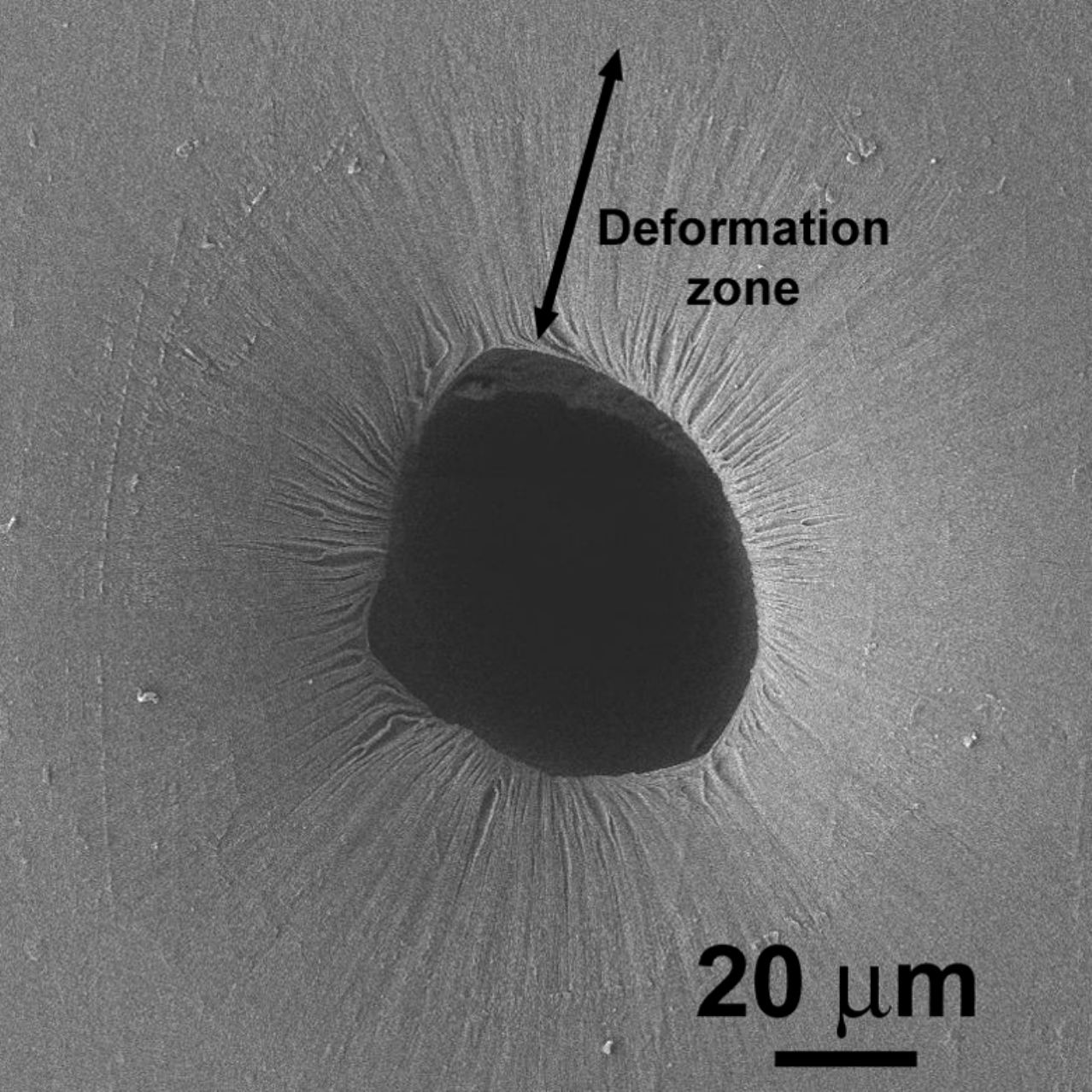}}
\subfigure[]{\label{SEM1Trans}\includegraphics[width=5.5cm]{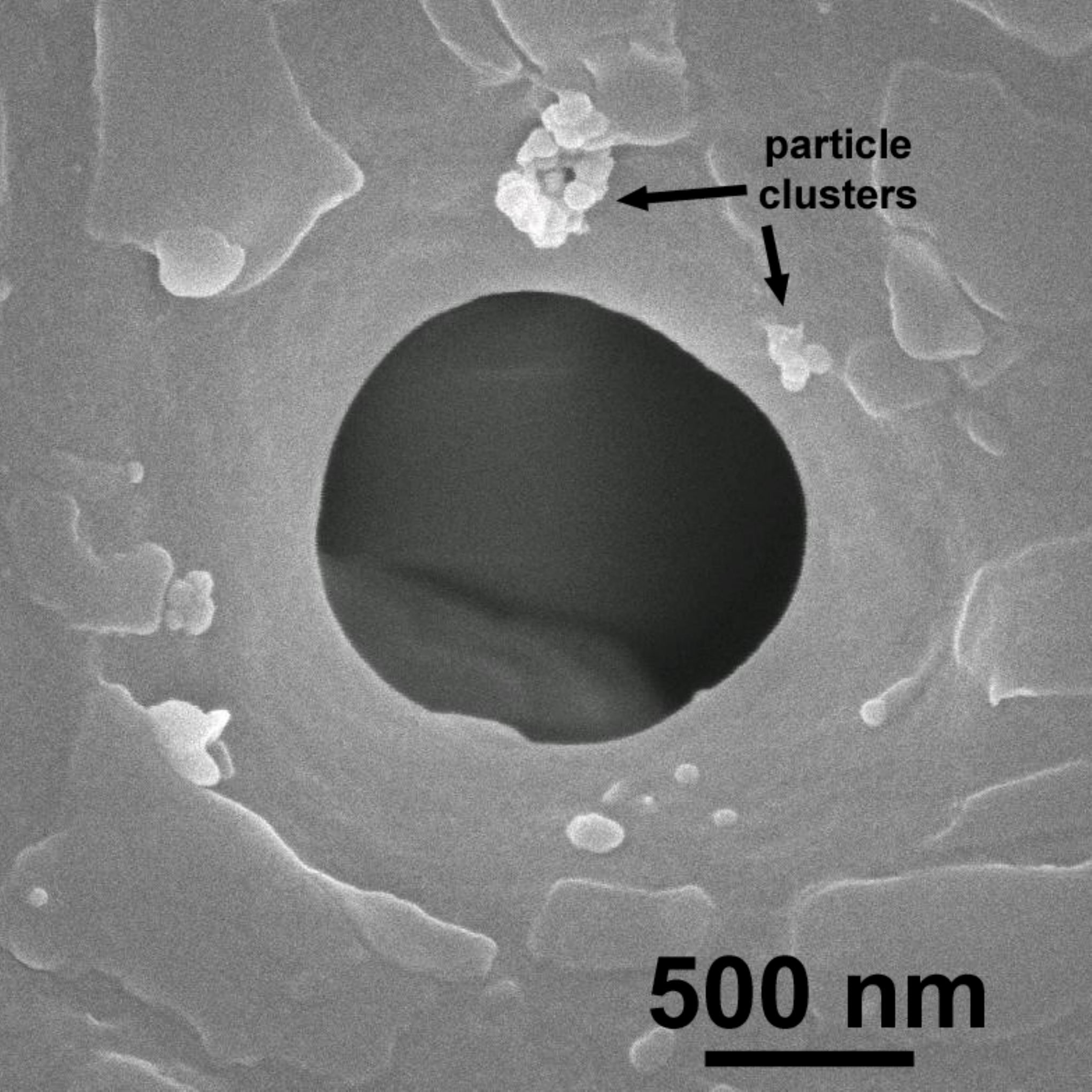}}
\subfigure[]{\label{SEM1Bore}\includegraphics[width=5.5cm]{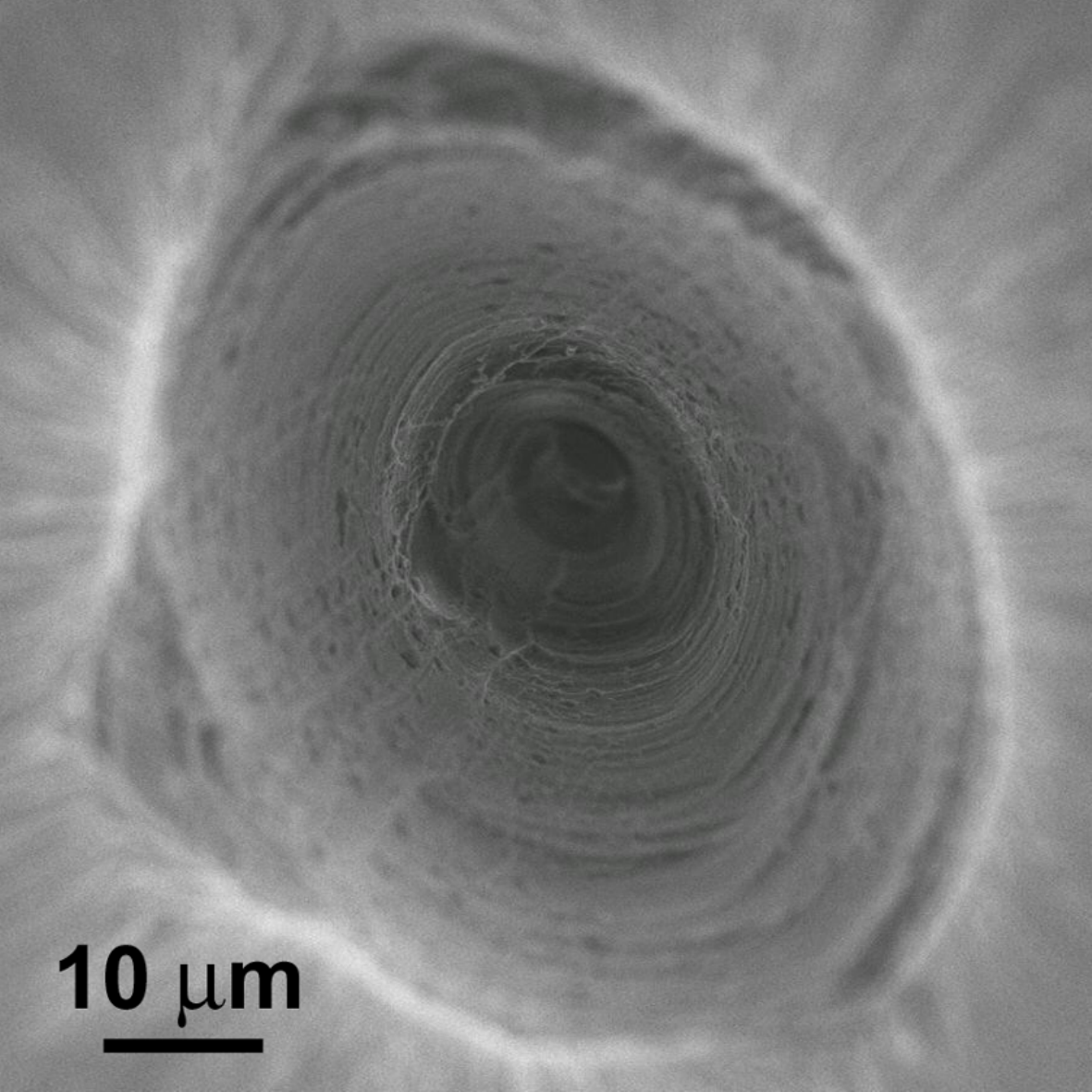}}
\end{center}
\caption{SEM imaging of (a) the cis- and (b) the trans- pore surfaces at $X=50$~mm. (c) Imaging of the bore of the pore, as viewed from the cis- side. Surfaces were coated prior to imaging, with the membrane held in a stretched position.}
\label{SEM1}       
\end{figure}

SEM images of the surfaces of a typical pore are shown in figure~\ref{SEM1}. The image of the stretched cis- entrance (figure~\ref{SEM1Cis}) shows that there is a zone of deformation near the pore, characterised by radial striations in the surface texture. This deformation is caused by inelastic overextension of the TPU during fabrication. Both cis- and trans- (figure~\ref{SEM1Trans}) pore entrances are reasonably circular; the fabrication process has developed since earlier SEM work \cite{738,660}, so that circular pore sections are more commonly obtained. 

Near the trans- opening, some particles are visibly attached to the TPU surface, including small aggregations of such particles. These are commercial carboxylate-modified polystyrene spheres, previously used in experiments with this membrane. Particle aggregation and adhesion to the TPU surface has occurred during the experiments. Usually such aggregation is undesirable, and the colloidal dispersion is expected to be stable in the electrolyte used. TPU surface potentials are important for electrophoretic transport as well as particle adhesion, and will be the subject of ongoing research.  

It is also possible to image the bore of the pore using SEM (figure~\ref{SEM1Bore}). At length scales greater than a few microns, the bore has a smooth, tapered internal profile. Small-scale ridges observed on the pore walls are a consistent feature of bore images. The observed spacing of the bore ridges (1-2~$\mu$m) is consistent with the expected spacing of TPU hard and soft chain regions at failure, for example near 500\% elongation \cite{738, 648}.

\begin{figure}
\begin{center}
\subfigure[]{\label{SEM2Cis40}\includegraphics[width=5.5cm]{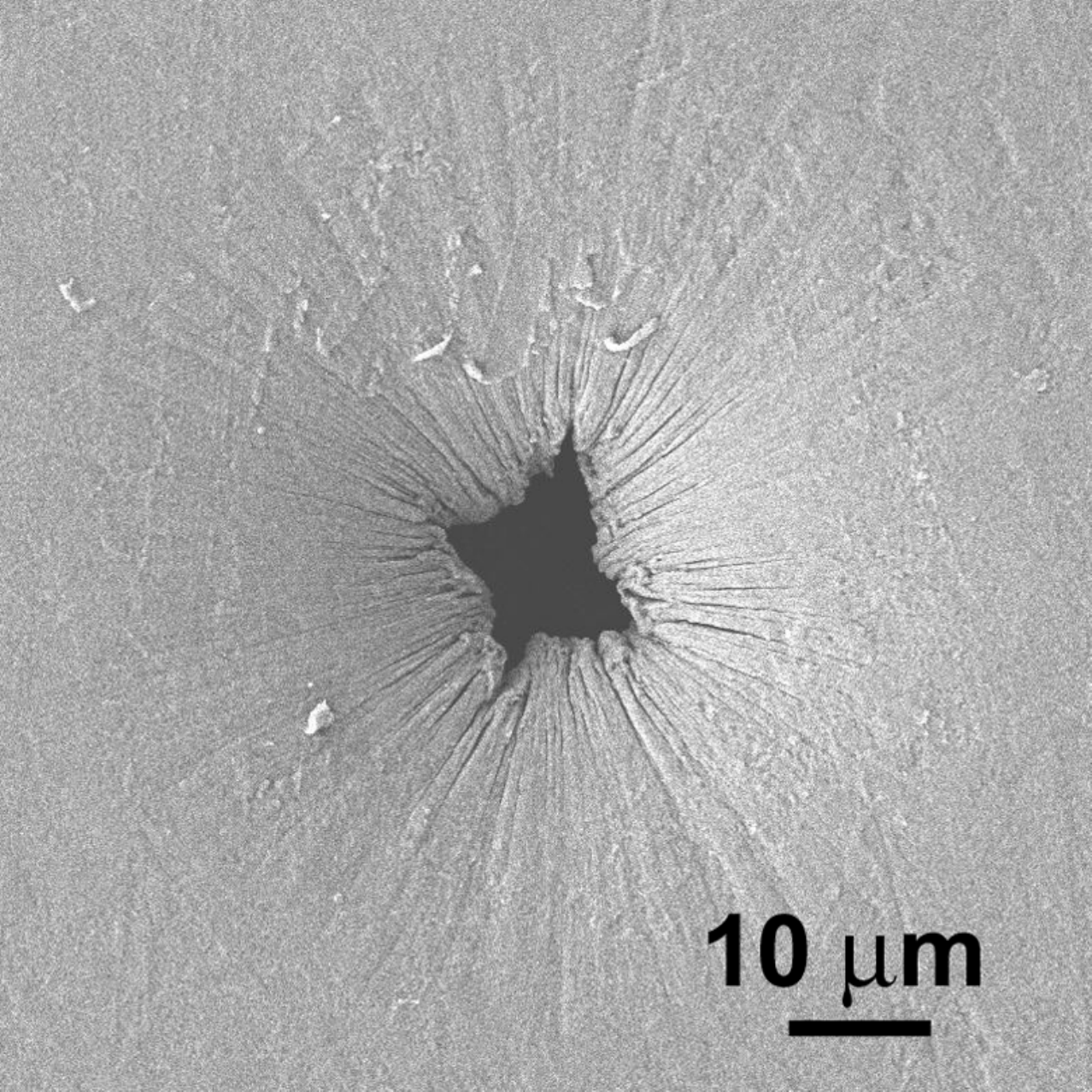}}
\subfigure[]{\label{SEM2Cis45}\includegraphics[width=5.5cm]{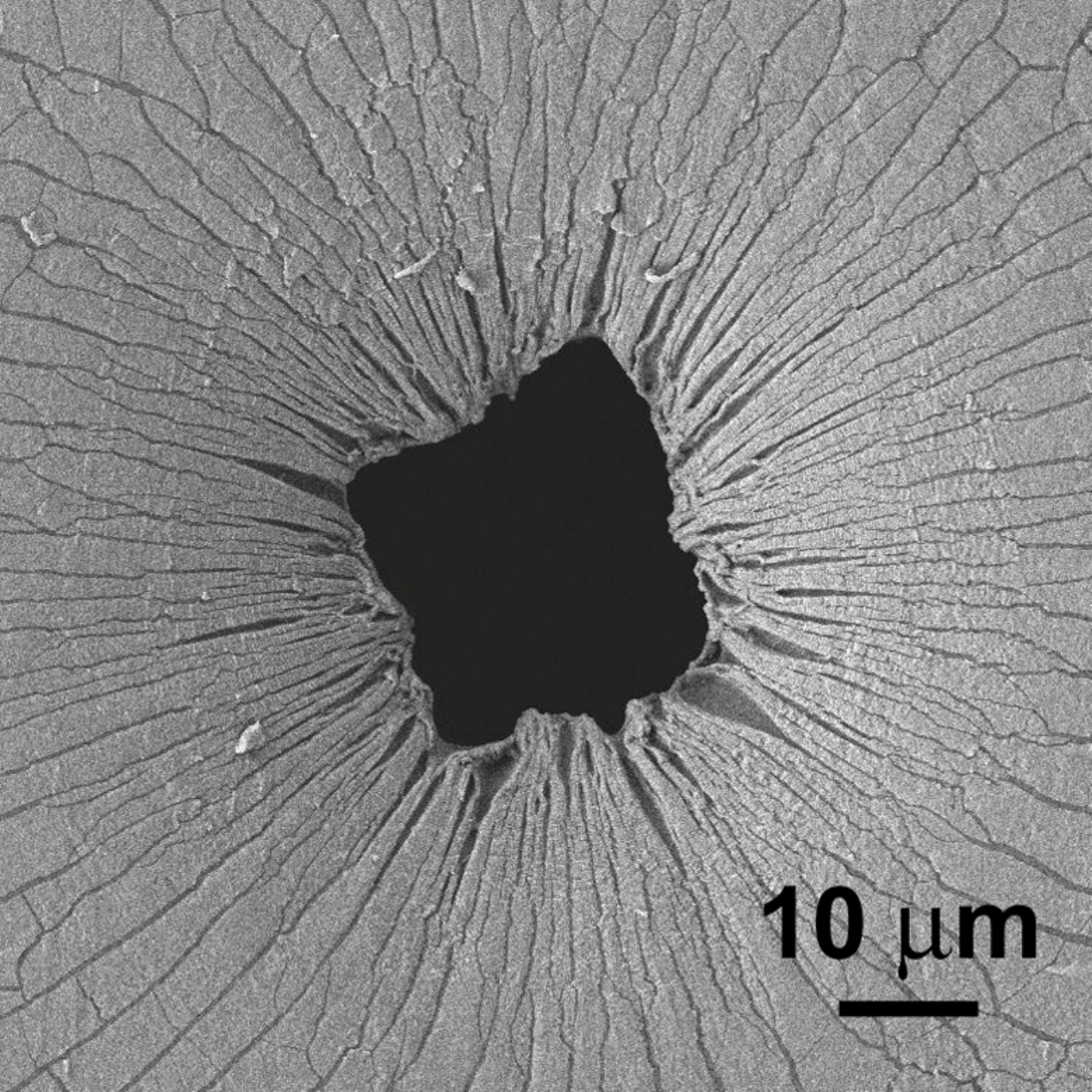}}
\subfigure[]{\label{SEM2Cis50}\includegraphics[width=5.5cm]{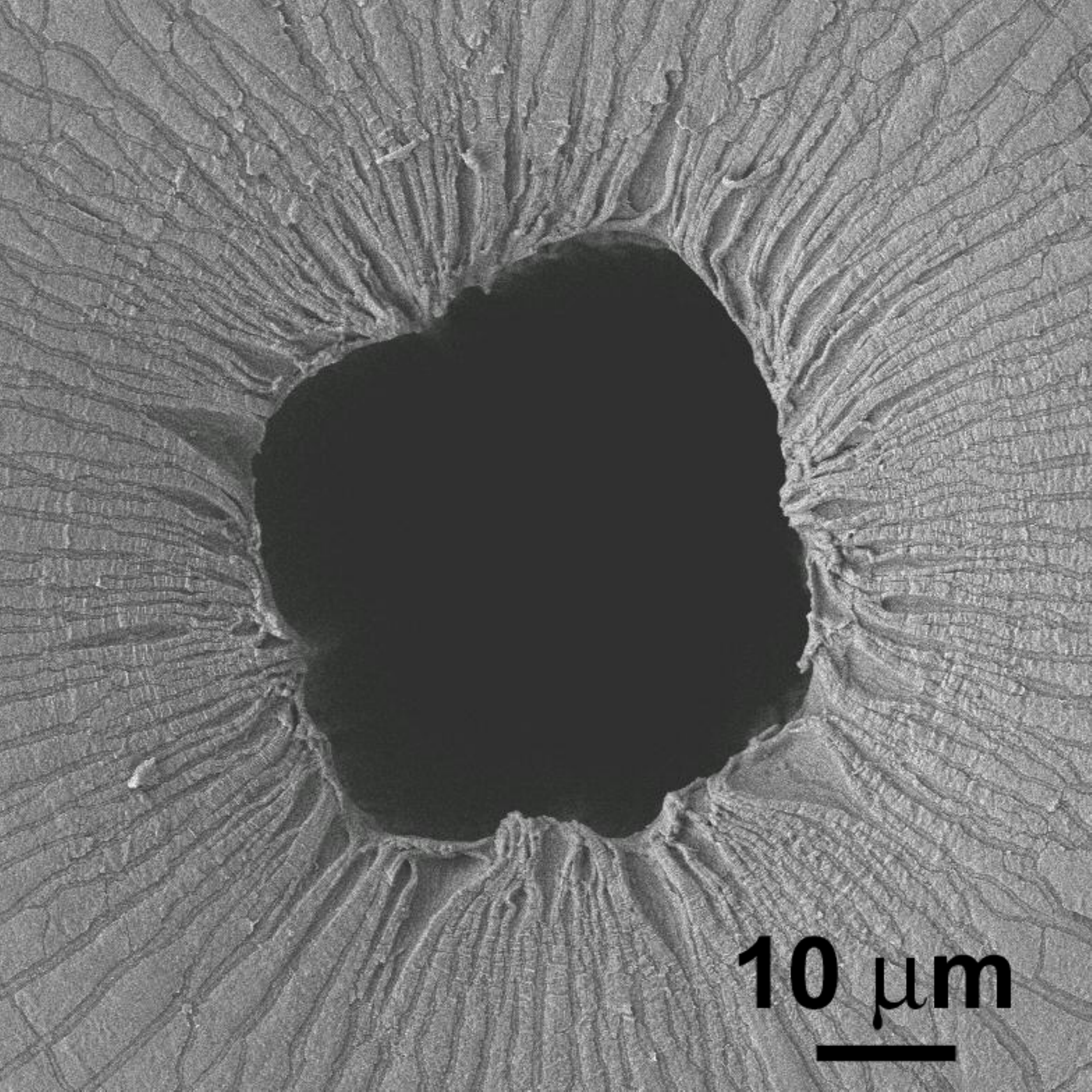}}
\subfigure[]{\label{SEM2Trans}\includegraphics[width=5.5cm]{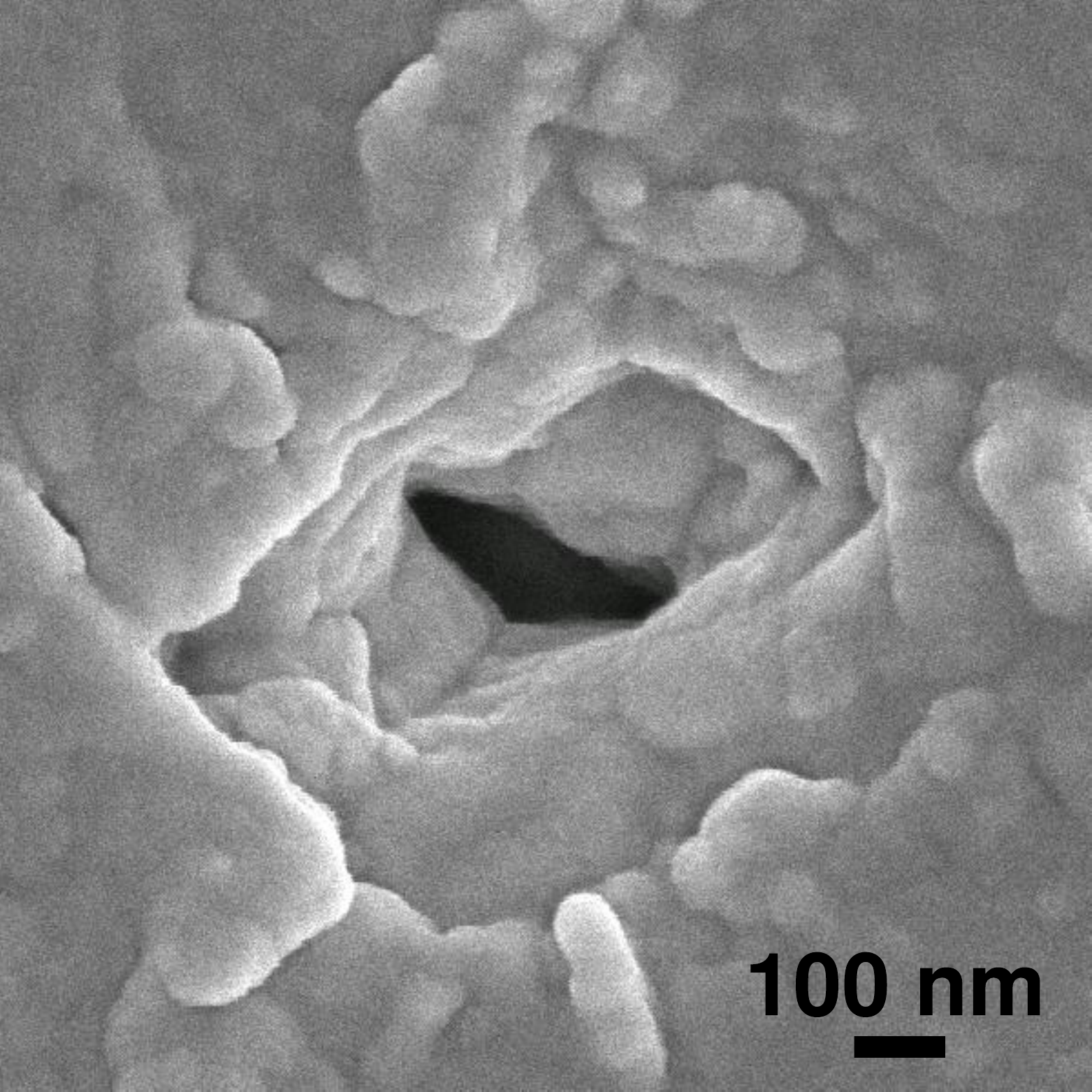}}
\end{center}
\caption{SEM imaging of the cis- surface of a pore (a) unstretched at $X_0=40$~mm (b) at $X=45$~mm, and (c) at $X=50$~mm. The trans- surface of the same pore at $X=50$~mm is shown in (d).
}
\label{SEM2cis}       
\end{figure}

The cis- surface of a second membrane (figure~\ref{SEM2cis}) shows the important role of the near-pore deformed area during stretching. At low stretch, the pore shape is asymmetric (star-shaped) with deep radial striations: material that was overextended during pore fabrication has been buckled and folded. When the membrane is stretched, overextended material can unfold and extend, so the profile becomes more azimuthally symmetric (circular) and the radial striations are shallower. 

This bunching behaviour has a significant, complicating effect on measurements of the pore size as a function of applied strain. For example, measurements from figure~\ref{SEM2cis} (see table~\ref{Strain}) show that the change in pore size is highly responsive over the stretching range studied, being greater than and non-linearly related to $\alpha$. These measurements represent the mean strain between four pairs of fiducial points in the images. The pore opening in figure~\ref{SEM2cis} is smaller than other pores imaged in the present work, so the effect of near-pore deformation is perhaps exacerbated. The key observations consistently arising from strain measurements such as these are (i) highly responsive pore size and (ii) variability between different cruciforms.

The trans- surface of the same pore (figure~\ref{SEM2Trans}) did not appear to change in size with stretch, within scaling error. Stretching of the trans- opening is less responsive than for the cis- opening. The opening is smaller than the corresponding image in figure~\ref{SEM1Trans}, and is not regularly shaped. Circular surface openings are difficult to reproduce when the size of the pore is less than the inherent spacing of hard and soft regions in the molecular structure of the TPU.

Away from the pore itself, cracking patterns are observed on the membrane surface in figure~\ref{SEM2cis}. These patterns are an artefact of the imaging process; the metallic Au/Pd coating cracks when the membrane is stretched. A coating was first applied prior to imaging at $X_0$~mm, so no cracking appears in figure~\ref{SEM2Cis40}. Further coatings were applied prior to subsequent images, so two levels of cracking can be observed on close inspection of figure~\ref{SEM2Cis50}.

\begin{table}
\caption{\label{Strain}Pore size strain measurements from imaging, all measured relative to images at $X_0 = 40$~mm ($\alpha=0$).}
\begin{tabular*}{\textwidth}{@{}l*{15}{@{\extracolsep{0pt plus
12pt}}l}}
\br
Figure&Surface&$X$&$\alpha$&Strain\\
\mr
SEM&&&&\\
\ref{SEM2Cis45}&Cis&45.0&0.125&1.1~$\pm$~0.2\\
\ref{SEM2Cis50}&Cis&50.0&0.25&2.5~$\pm$~0.4\\
\mr
Confocal&&&\\
\ref{Conf45}&Cis&44.9&0.123&0.14~$\pm$~0.01\\
\ref{Conf50}&Cis&50.0&0.25&0.58~$\pm$~0.01\\
\ref{Conf45}&Trans&44.9&0.123&0.09~$\pm$~0.09\\
\ref{Conf50}&Trans&50.0&0.25&0.18~$\pm$~0.09\\
\mr
\br
\end{tabular*}
\end{table}
 
\subsection{Confocal microscopy} 

\begin{figure}
\begin{center}
\subfigure[]{\label{Conf3D}\includegraphics[width=8cm]{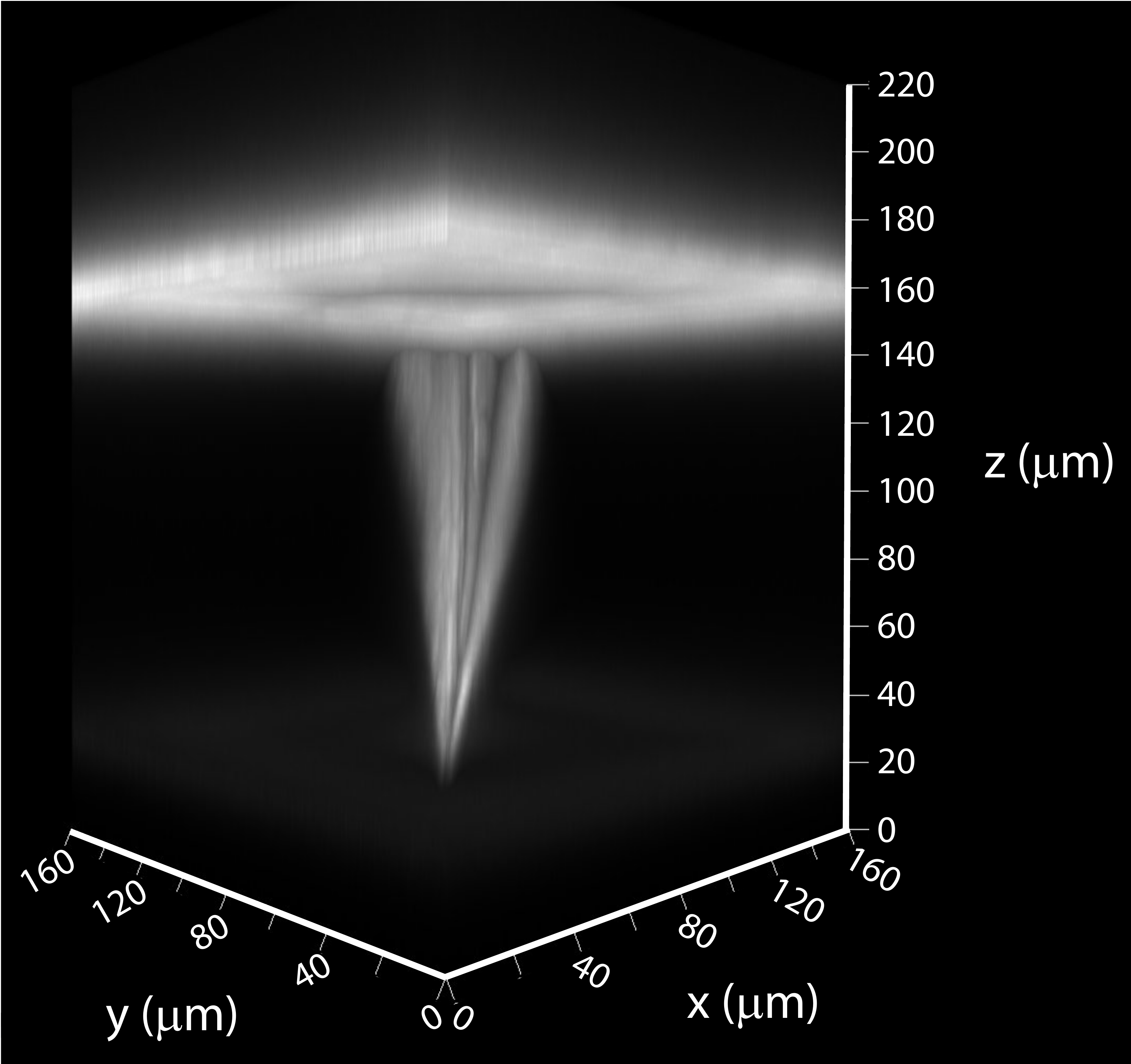}}
\newline
\subfigure[]{\label{Conf40}\includegraphics[width=4cm]{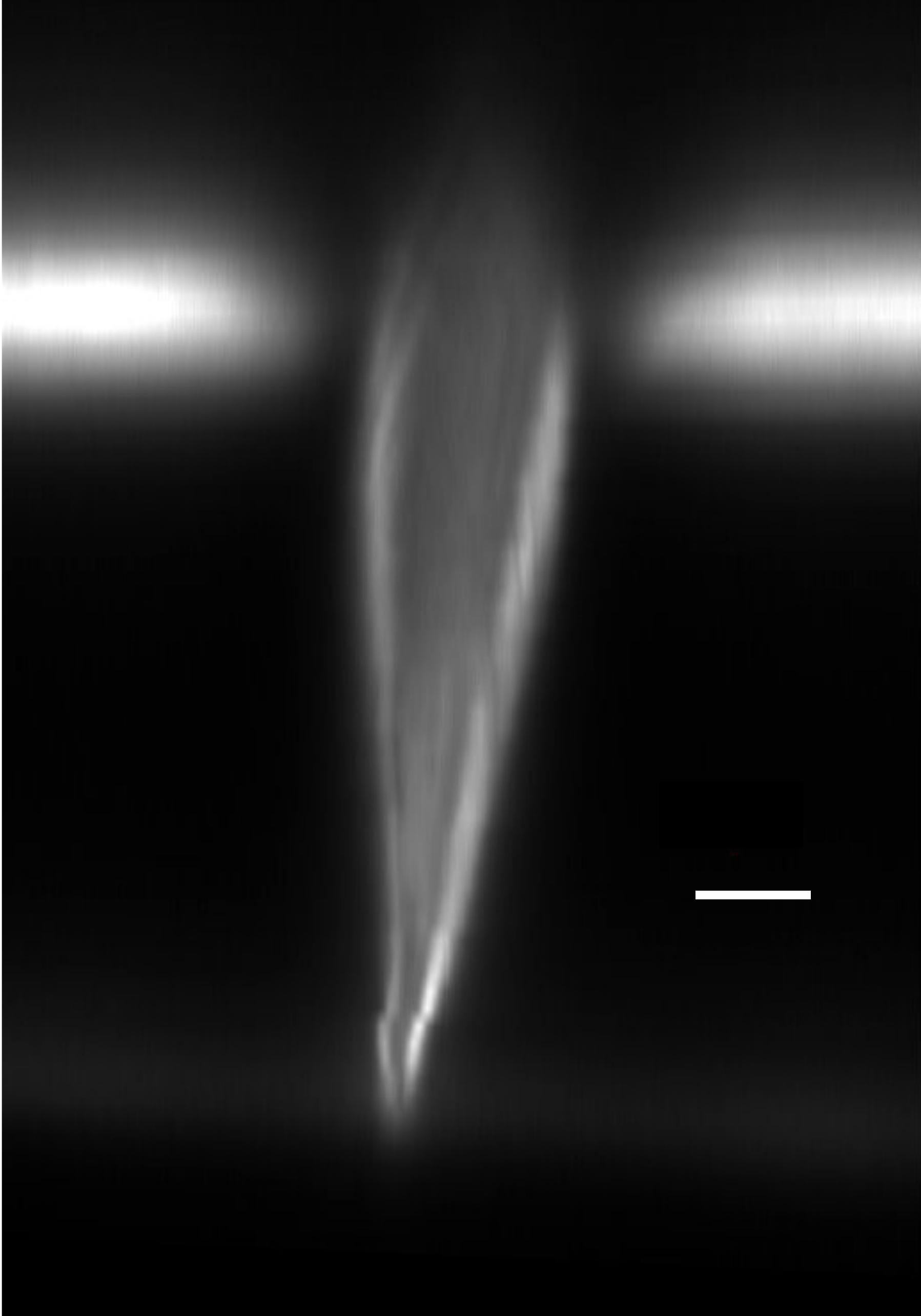}}
\subfigure[]{\label{Conf45}\includegraphics[width=4cm]{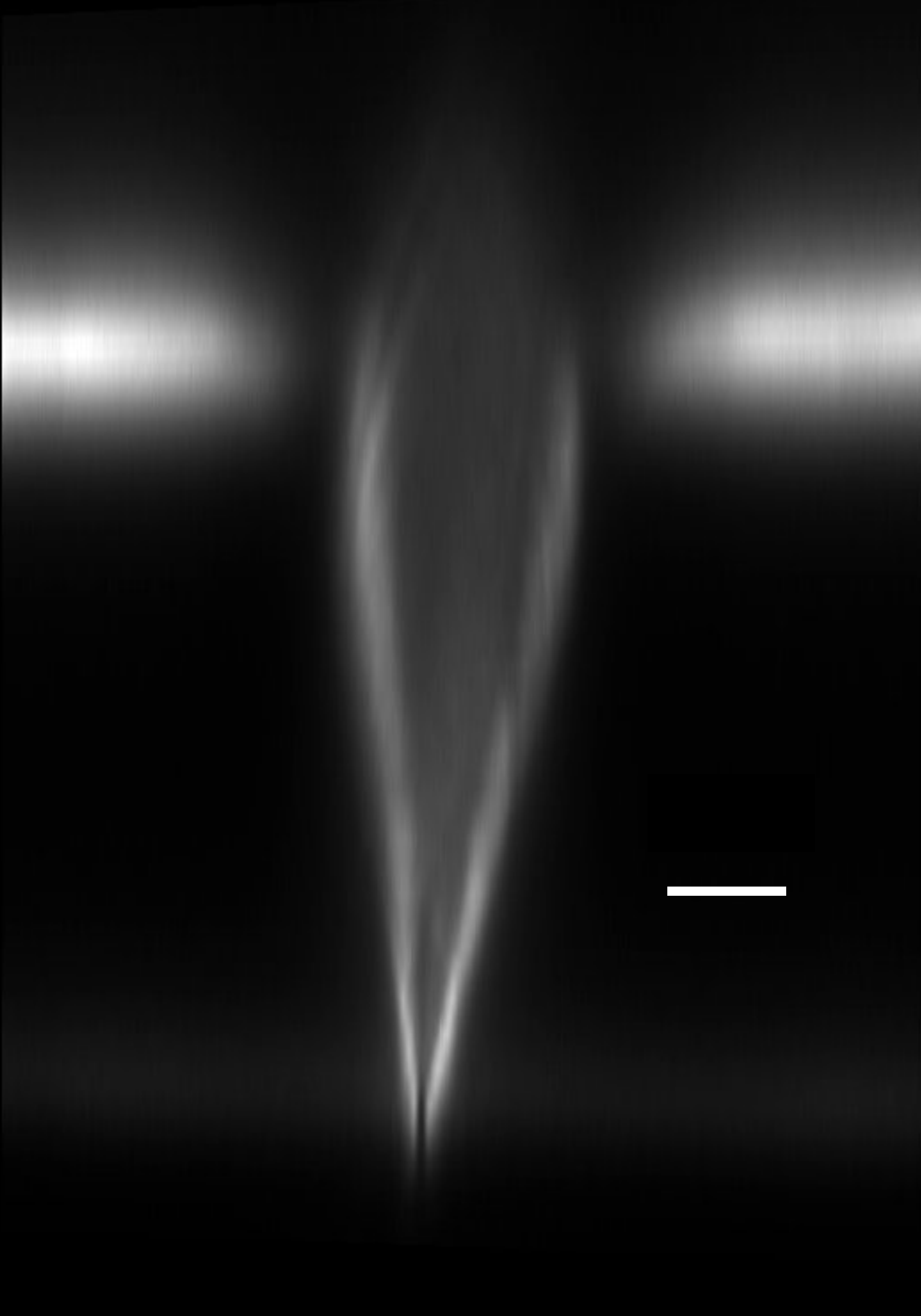}}
\subfigure[]{\label{Conf50}\includegraphics[width=4cm]{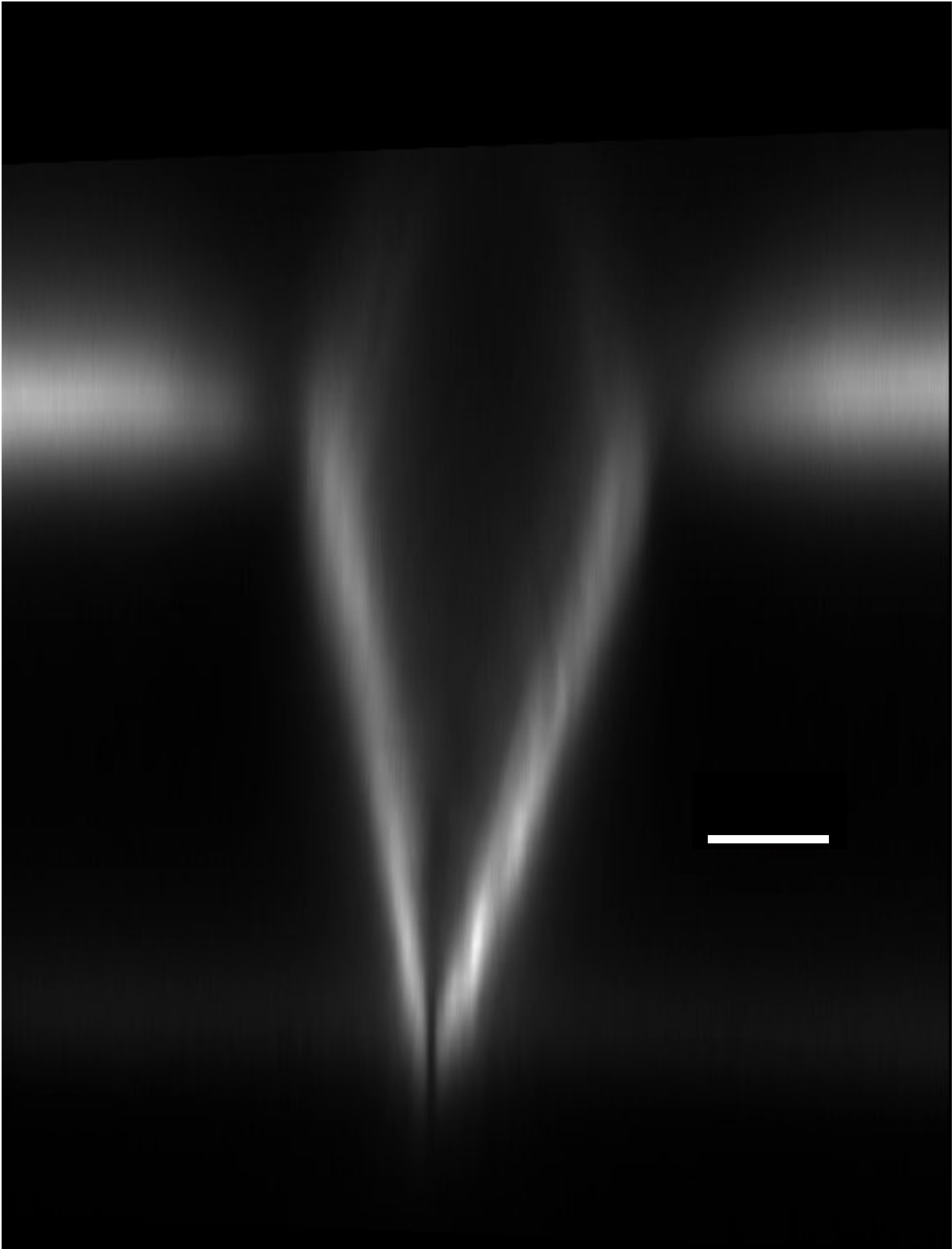}}
\end{center}
\caption{Confocal microscopy of a nanopore showing (a) a 3D projection, reconstructed from z-stack images, of the pore shape at $X=44.9$~mm, and cross-sectional images produced by clipping the 3D image (b) unstretched at $X_0$, (c) at $X=44.9$~mm, and (d) at $X=50$~mm. Scale bars in each image represent 20~$\mu$m. In (b), the distortion near the trans- surface is a result of a small movement of the microscope stage between capture of image slices.}
\label{Conf}       
\end{figure}

Figure~\ref{Conf3D} shows a 3D image, while figures~\ref{Conf40} to \ref{Conf50} are cross-sectional images of the pore at three stretch settings, showing changes in pore geometry with increased stretch. Penetration of the dye solution into the polymer membrane is illustrated in these images. The cis- surface of the membrane shows more fluorescence than the trans- surface because it was in constant contact with the dye solution as it was run through the pore, whereas the trans- surface was initially in contact only with standard buffer. The region adjacent to the pore on the cis- surface did not absorb the dye as readily as other surfaces, and appears dark in the images. This region corresponds to the deformation zone observed in SEM images, providing further evidence that the properties of this material in this region have been significantly altered during the fabrication process.

The width of the pore at the cis- and trans- surfaces has been measured from the cross-sectional images, with results shown in table~\ref{Strain}. The pore strain at the cis- surface is greater than and non-linearly related to $\alpha$, although to a far lesser extent than observed in figure~\ref{SEM2cis}. These pore strain measurements explain the difficulty in fitting current data to simple models, even when the near-pore strain is well known \cite{660,771}.

Measurement of the trans- opening strain is comparable to $\alpha$. It is encouraging that a size change is observed, given that the constriction of the pore at or near the trans- surface opening is the most important and sensitive section of the pore for applications. The observation that trans- pore strain is comparable to the applied macroscopic strain also suggests that deformed material adjacent to the pore does not affect stretching. 

With increasing stretch the membrane became thinner, reducing the pore length. This effect was studied by measuring the distance between the outer surfaces of the membrane. The pore thinned by 4\% and 18\% at $X=44.9$~mm and $X=50$~mm respectively, whereas the degree of thinning predicted by the simple linear elastic model \cite{771} is 25\% and 50\%. This kind of discrepancy can lead to an error of, for example, 40\% at $X=50$~mm for the value of $l$ in analysis of (\ref{eq:terms}) or (\ref{eq:con}). The large underestimate is consistent with the presence of folded material adjacent to the pore, but could also reflect the full viscoelastic properties of pristine TPU. 

The resistance of the pore in figure~\ref{Conf} was measured while filled with SEB ($\rho=$~0.86~$\Omega$~m). In the cases of figures~\ref{Conf45} and \ref{Conf50}, data is available for direct comparison between the measured resistance, and the value calculated using (\ref{eq:con}). At $X=44.9$~mm, the measurement was 5.4~$\pm$~0.1~M$\Omega$ and the calculated value was 4.7~$\pm$~1.2~M$\Omega$; at $X=50$~mm the measurement was 2.2~$\pm$~0.1~M$\Omega$ and the calculated value was 2.6~$\pm$~0.6~M$\Omega$. Therefore, the conical-pore model with dominant electrophoresis (\ref{eq:con}) is very consistent with the experimental measurement, even when the pore was stretched. However, use of (\ref{eq:con}) to determine $a$ requires a measurement of the pore resistance along with knowledge of $b$ and $l$ at a given macroscopic strain. Quantitatively determining the pore geometry without using time-consuming techniques remains a challenge. Future development of the analytical model \cite{771} for pore tuning should incorporate (in order of importance) an overextended area of material near the pore, viscoelastic material properties, and pore cross-sections which are not linear cones. A better understanding of the mechanical tuning process will enable more rigorous interpretation of blockade rate measurements.

\section{\label{Darbytext}Blockade Rate Measurements}

\subsection{\label{presstext}Concentration of Particles in Pressure-Driven Flow}

\begin{figure}
\begin{center}
\subfigure[]{\label{PDFFvsP}\includegraphics[width=8cm]{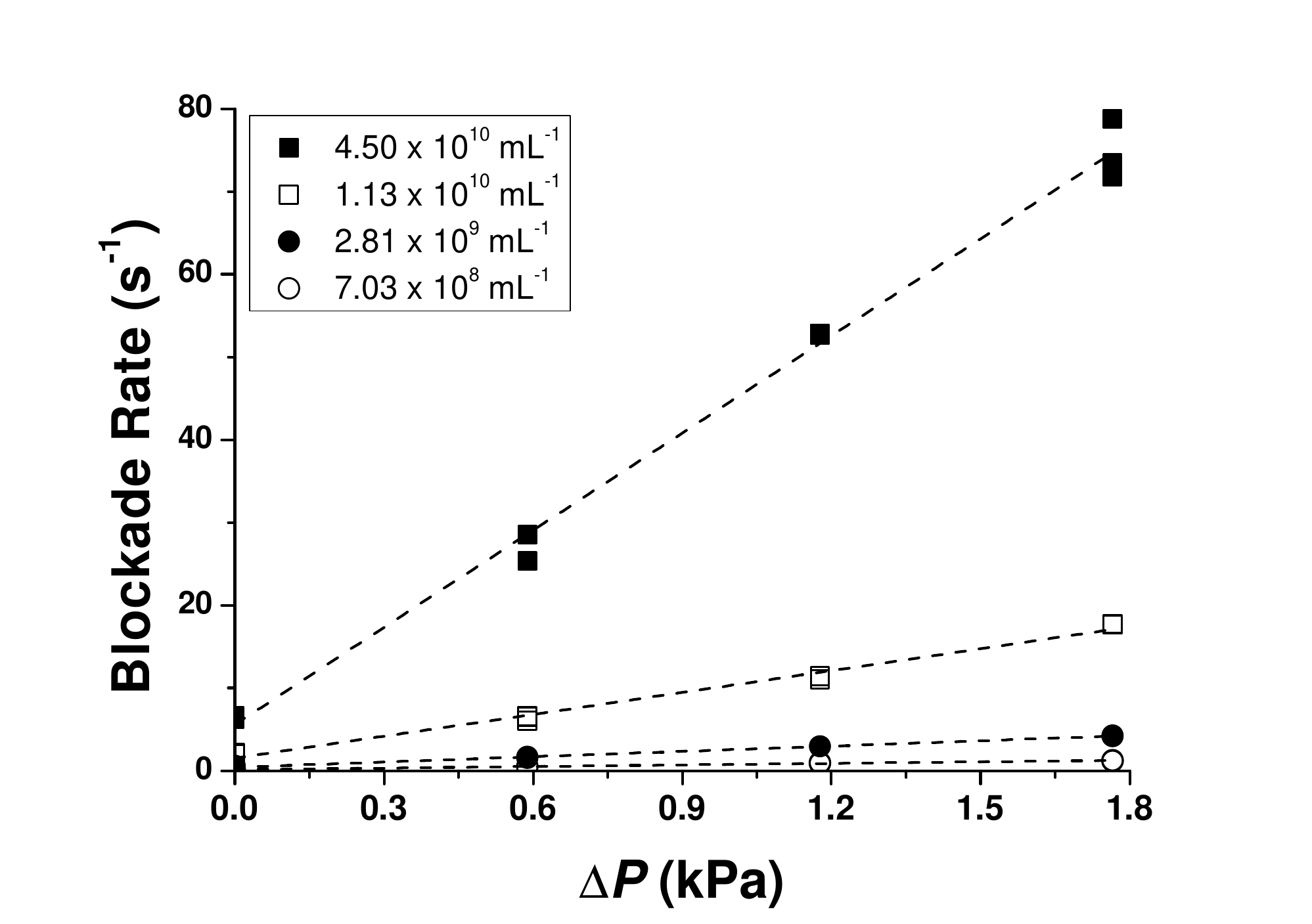}}
\newline
\subfigure[]{\label{PDFGvsC}\includegraphics[width=5.5cm]{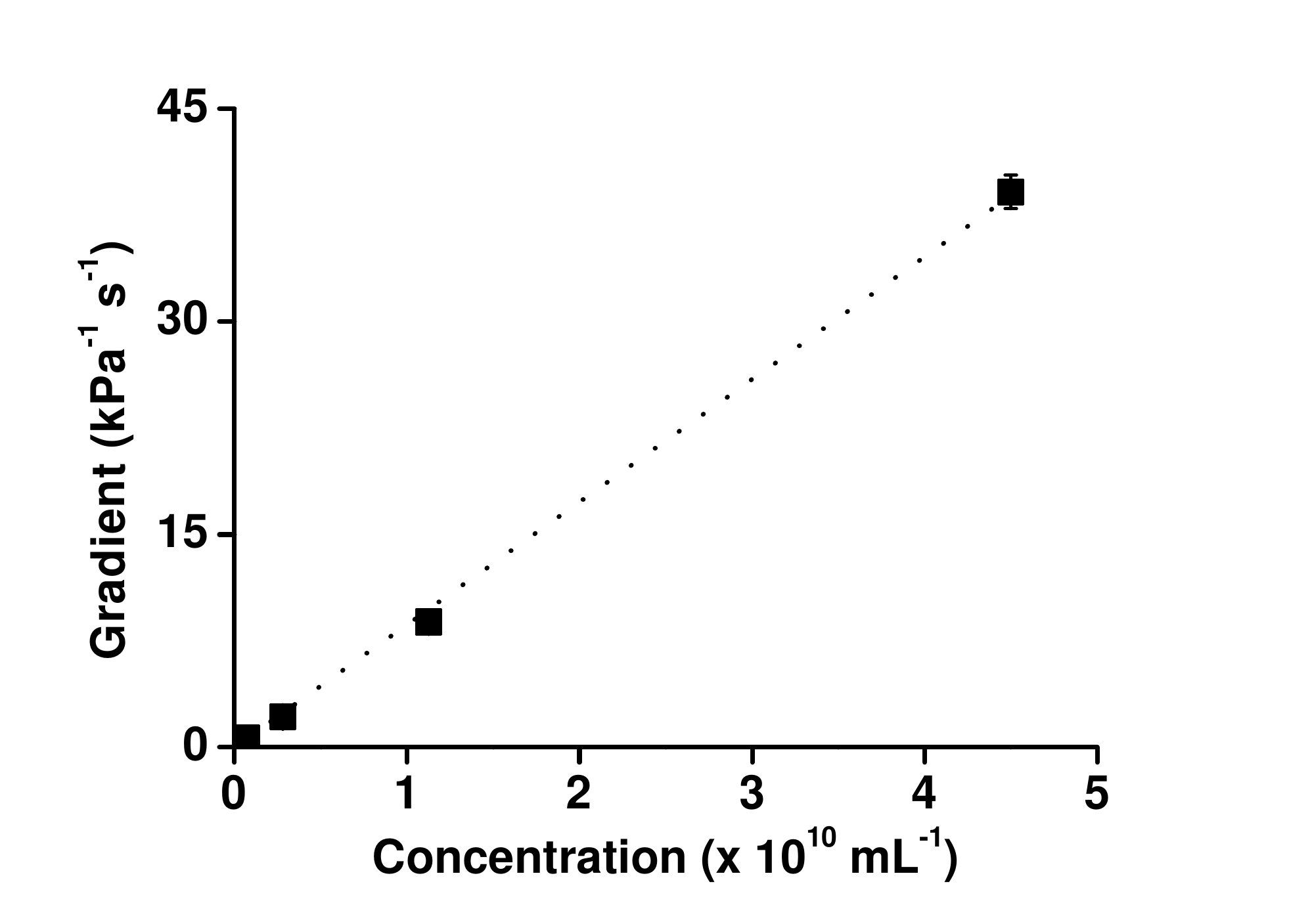}}
\subfigure[]{\label{PDFIvsC}\includegraphics[width=5.5cm]{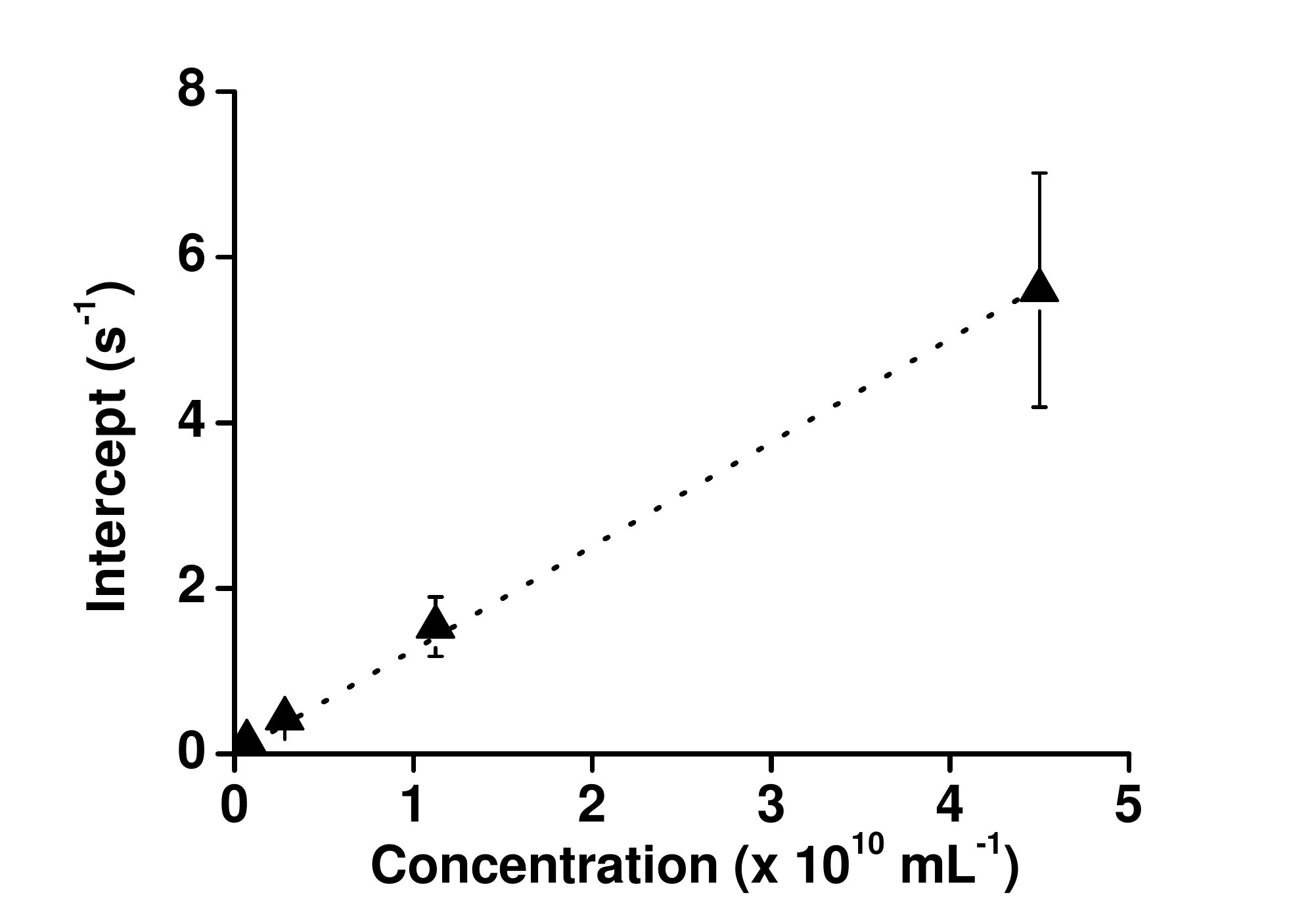}}
\end{center}
\caption{Blockade rate data for pressure-driven flow measurements. (a) Blockade rate as a function of applied pressure for four different particle concentrations. In each case, linear plots to the data ($R^2 \geq 0.985$) are shown. In (b) and (c) respectively, the gradients and intercepts from plot (a) are plotted as a function of particle concentration. Linear trends in (b) and (c) ($R^2 \geq 0.999$) are forced through the origin, and error bars represent standard errors calculated during fitting to plot (a). The ionic current record for each data point lasted at least one minute, and measurements at the same pressure and concentration were not recorded subsequently to each other.}
\label{PDF}       
\end{figure}

Experiments were carried out to test the dependence of blockade rate on concentration, with variable pressure-driven flow applied. Blockade rates measured at four different pressures and four different particle concentrations are shown in figure~\ref{PDF}. At each concentration, the blockade rate is a linear function of pressure (figure~\ref{PDFFvsP}). This linear functionality is consistent with the theoretical model (\ref{eq:terms}) for the particle flux with an applied pressure. Extending that same theory, we can predict that the blockade rate should be independent of particle type, including size and surface charge variations; a similar linear trend has previously been observed for a variety of particles \cite{882}.

A linear dependence was also found when the gradients and intercepts from figure~\ref{PDFFvsP} were plotted as a function of particle concentration, in figures~\ref{PDFGvsC} and \ref{PDFIvsC} respectively. From the gradient of figure~\ref{PDFGvsC} (8.64~x~10$^{-19}$~Pa$^{-1}$~s$^{-1}$~m$^3$), the effective pore size was calculated from (\ref{eq:terms}) to be $a_0=790 \pm 110$~nm. This calculation represents an advance on a similar method used previously \cite{882}, because uncertainty relating to the particle concentration has been reduced. The value of $a$ derived using the baseline current through the pore (calculated using (\ref{eq:con}); 33~nA at 0.3~V applied potential) is $330 \pm 110$~nm. Both values are notably smaller than the range previously calculated using similar methods for blockades caused by 200~nm spheres (1.1 - 2.4~$\mu$m \cite{882}), reflecting that the signal-to-noise for 100~nm particles is optimised for a smaller pore size setting.  

Assuming that the quoted concentration of as-received particles was accurate, the discrepancies and uncertainties in the two pore radius values are dominated by three geometric assumptions. In order of importance, the first assumption is that the pore profile takes a regular shape such as a cylinder or a cone. The value extracted from current measurements is smaller because that method calculates the smaller pore size of a tapering cone ($a$), rather than an effective average ($a_0$) down the length of a pore which is assumed cylindrical. Secondly, the value of the larger pore size $b$ is taken to be $15\pm5$~$\mu$m (at rest) on the basis of the history of observations of TNs using SEM. The large uncertainty reflects variation from pore to pore. Thirdly, the membrane thickness and value of $b$ are assumed to scale as if the material was linear and elastic, using the simple approach described in \cite{771}; errors due to this calculation were discussed in Section~\ref{2.2}.

The vertical axis intercepts ($\Delta P = 0$) in figure~\ref{PDFFvsP} are relatively small when compared with the blockade frequencies measured over the range of applied pressures; pressure-driven flow indeed dominates the experiments. The intercepts are a linear function of concentration (figure~\ref{PDFIvsC}), another feature of the data which is consistent with (\ref{eq:terms}). According to that analysis, the intercept is equal to the sum of electrophoretic and electro-osmotic contributions to transport, both of which are proportional to concentration. These two contributions cannot be distinguished using the present measurements, but could be studied by varying the particle type, applied potential, or (as in the following Section) the properties of the surfaces and the electrolyte. 

Future experiments will directly test the hypothesis that blockade rate is independent of particle type in pressure-driven flow. Ateempts will be made to establish a simple calibration procedure for fast, precise measurement of unknown particle concentrations in the tens to hundreds of nanometres range. Such a technique is limited by the accuracy of the calibration sample concentration - as recorded by a commercial supplier, for example.

\subsection{\label{zeta}Surface Charge Dependence of Electrophoretic Transport}

A second set of experiments aimed to study the effect of electrophoretic transport of particles through the pore, measured by particle blockade rate. The electrophoretic particle flux in (\ref{eq:terms}) is determined by particle mobility, which is greatest when the effective charge on the particle is large (\ref{eq:zeta}). Depending on the surface functionality of the particles, the degree of charge changes as a function of the pH of the solution. The effects of pH on particle charge depend on the pKa of the functional groups and strongly adsorbed counter-ions on the particle surface. Acidic and basic functional groups impart a negative or positive charge when deprotonated or protonated, respectively. A TN was used to study the trend between blockade rate and pH for two silica particle sets: 200~nm and 500~nm diameter particles with carboxylic acid and amine surface functionalities respectively. The trend is directly compared with independent zeta potential measurements using the conventional laser doppler velocimetry technique, which calculates the zeta potential of a particle dispersion from the measured particle mobility using (\ref{eq:zeta2}).

\begin{figure}
\begin{center}
\subfigure[]{\label{DarbyA}\includegraphics[width=8cm]{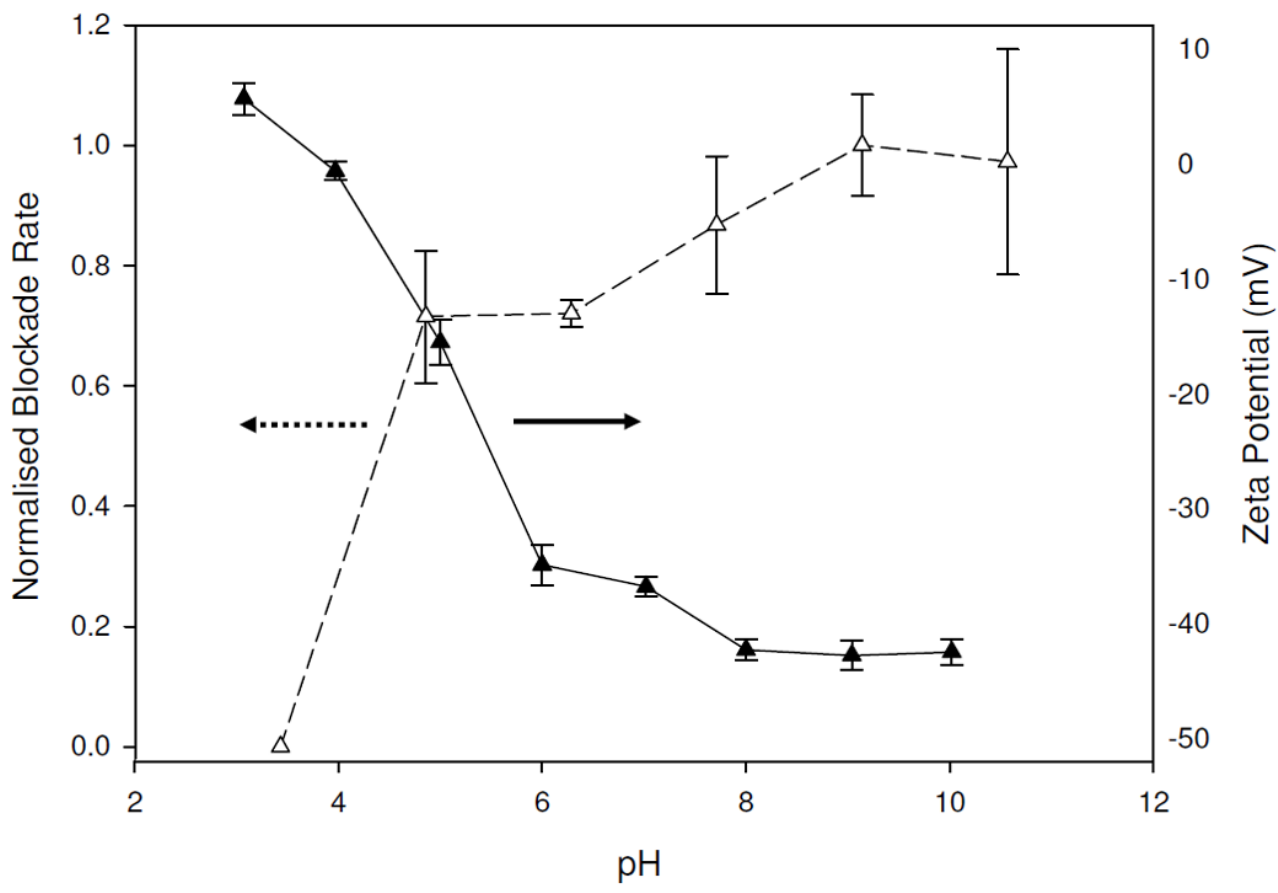}}
\subfigure[]{\label{DarbyB}\includegraphics[width=8cm]{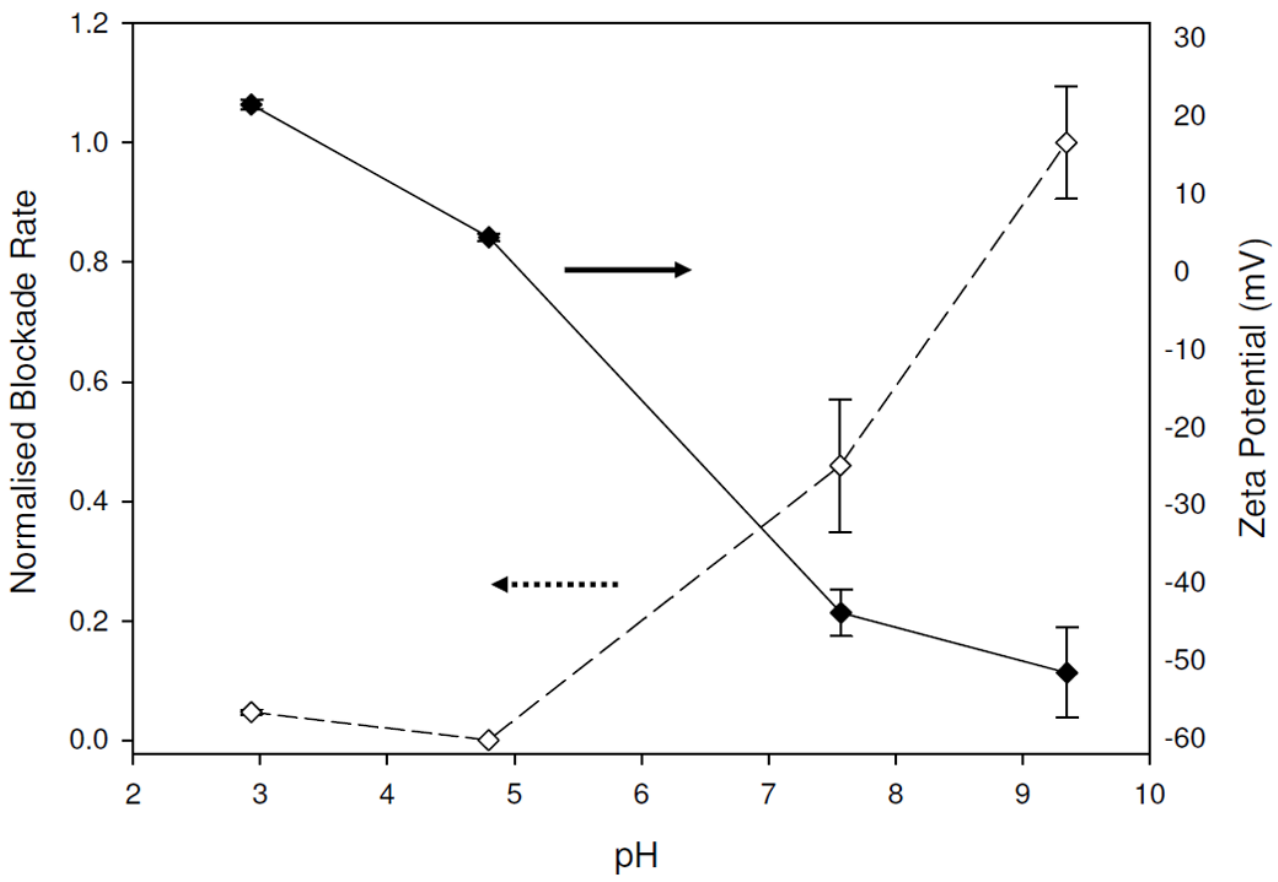}}
\end{center}
\caption{Effects of particle charge on rate of passage through a size-tuneable, electrophoretically driven nanopore.  (a) Zeta potential ($\blacktriangle$) and nanopore particle blockade rate ($\opentriangle$) of 500~nm carboxylic acid modified silica particles. (b) Zeta potential ($\blacklozenge$) and nanopore particle blockade rate ($\opendiamond$) of 200~nm amine modified silica particles. All data points represent the average and standard deviation of three individual runs, each consisting of a one minute ionic current record.}
\label{Darby}       
\end{figure}

Figure~\ref{Darby} shows that the blockade rate increased with increasing zeta potential magnitude for both particle sets. The highest rate corresponded to the highest pH values tested, where both sets of particles were most negatively charged (zeta potential $<-40$~mV). As the pH of the solution was reduced, the zeta potential of both particle sets decreased until they reached their respective isoelectric points, being approximately at pH 4 and 5.5 for the carboxylic acid and amine modified silica particles, respectively. A similar decrease in rate was observed for both particle sets until no particles were seen to traverse the pore near the isoelectric points. As the particles passed through their isoelectric point and switched charge, the polarity of the nanopore cell voltage was similarly switched to continue to measure blockade rate. This is most clearly demonstrated for the amine functionalised silica particles in figure~\ref{DarbyB} for which the zeta potential varied between 20~mV at pH~3 to -50~mV above pH~8. 

Interestingly, particles with zeta potentials between $\pm$10~mV, near the isoelectric point, were found to be insufficiently charged to be driven through the pore. Furthermore, under these conditions the particle suspensions became unstable, giving rise to particle aggregation and blocking of the nanopore. This corresponds with general zeta potential theory relating to colloid stability where typically, zeta potentials of magnitude greater than 30~mV are considered to be strongly charged, and therefore more likely to form stable particle suspensions. Below 14~mV magnitude, attractive forces between particles often become greater than the electrostatic repulsive forces, causing the particles to aggregate \cite{891}. 

It is hypothesized that the effective charge (and therefore, zeta potential) of individual particles can be measured with TNs, for example using (\ref{eq:terms}) under the assumption that electro-osmotic transport is not significant. For quantitative measurements, further experimentation - similar to the work described for pressure-driven flow in Section~\ref{presstext} - is required to test assumptions and establish calibration techniques. 

\section{Conclusions}

Two experiments have demonstrated high potential of TN technology for characterising colloidal dispersions. The first experiment suggests that the concentration of an unknown specimen could be quickly determined, once a calibration using a particle set of known concentration had been carried out. The second experiment demonstrates a method for studying particle electrophoretic mobility at the level of individual particles. Both types of experiment provide quantitative blockade rate data that can be compared with Nernst-Planck theory for particle transport. The use of TNs to characterise colloidal charge, concentration and other interesting parameters in this manner is timely and efficient in comparison with existing techniques.

There is still progress to be made in the development of TNs, and this paper has highlighted some particularly important issues. Further imaging of pores, especially using the confocal technique, is required in order to obtain accurate geometrical parameters for analysis, and to understand the variability from pore-to-pore. The SEM images in particular have revealed a key requirement for any model describing pore tuning: inclusion of the inelastically deformed area adjacent to the pore on the cis- surface. The surface potential of TPU should be studied in order to reduce uncertainties relating to electro-osmotic flow calculations. Finally, the system for acquisition and processing of electronic signals should be further studied so that rigorous blockade pulse magnitude and duration data can also be applied in quantitative studies.

\ack

The authors thank Sylvia Zellhuber-McMillan (IZON) and Liz Girvan (Otago Centre for Electron Microscopy) for technical assistance with the SEM imaging, and Ben Glossop (IZON) for technical development of the qNano apparatus. This work was partly funded by the Australia New Zealand Biotechnology Partnership Fund and New Zealand's Foundation for Research, Science and Technology.


\begin{thebibliography}{39}
\expandafter\ifx\csname natexlab\endcsname\relax\def\natexlab#1{#1}\fi
\expandafter\ifx\csname bibnamefont\endcsname\relax
  \def\bibnamefont#1{#1}\fi
\expandafter\ifx\csname bibfnamefont\endcsname\relax
  \def\bibfnamefont#1{#1}\fi
\expandafter\ifx\csname citenamefont\endcsname\relax
  \def\citenamefont#1{#1}\fi
\expandafter\ifx\csname url\endcsname\relax
  \def\url#1{\texttt{#1}}\fi
\expandafter\ifx\csname urlprefix\endcsname\relax\def\urlprefix{URL }\fi
\providecommand{\bibinfo}[2]{#2}
\providecommand{\eprint}[2][]{\url{#2}}

\bibitem{562} Dekker C 2007 {\it Nat. Nanotechnol.} {\bf 2} 209--215

\bibitem{776} van Dorp S, Keyser U F, Dekker N H, Dekker C and Lemay S G 2009 {\it Nat. Phys.} {\bf 5} 347--351

\bibitem{611} Kasianowicz J J, Brandin E, Branton D and Deamer D W 1996 {\em Proc. Natl. Acad. Sci. USA} {\bf 93} 13770--13773
  
\bibitem{770} Clarke J, Wu H -C, Jayasinghe L, Patel A, Reid S and Bayley H 2009 {\it Nat. Nanotechnol.} {\bf 4} 265--270
  
\bibitem{570} Cockroft S L, Chu J, Amorin M and Ghadiri M R 2008 {\it J. Am. Chem. Soc.} {\bf 130} 818--820

\bibitem{572} Benner S, Chen R J A, Wilson N A, Abu-Shumays R, Hurt N, Lieberman K R, Deamer D W, Dunber W B and Akeson M 2007 {\it Nat. Nanotechnol.} {\bf 2} 718--724

\bibitem{895} Deamer D W and Branton D 2002 {\it Acc. Chem. Res.} {\bf 35} 817--825

\bibitem{589} Song M R, Hobaugh M R, Shustak C, Cheley S, Bayley H and Gouaux J E 1996 {\it Science} {\bf 274} 1859--1866

\bibitem{505} Li N, Yu S, Harrell C C and Martin C R 2004 {\it Anal. Chem.} {\bf 76} 2025--2030

\bibitem{530} Sexton L T, Horne L P and Martin C R 2007 {\it Mol. BioSyst.} {\bf 3} 667–-685

\bibitem{752} Ito T, Sun L and Crooks R M 2003 {\it Anal. Chem.} {\bf 75} 2399–-2406

\bibitem{580} Xue Y and Chen M 2006 {\it Nanotechnology} {\bf 17} 5216–-5223

\bibitem{881} Siwy Z S and Davenport M 2010 {\it Nat. Nanotechnol.} {\bf 5} 174–-175

\bibitem{778} Sun L and Crooks R M 2000 {\it J. Am. Chem. Soc.} {\bf 122} 12340--12345

\bibitem{877} DeBlois R W, Bean C P and Park R K A 1977 {\it J. Colloid. Interf. Sci.} {\bf 61} 323--335

\bibitem{767} Henriquez R R, Ito T, Sun L and Crooks R M 2004 {\it Analyst} {\bf 129} 478--482

\bibitem{884} Sachs F 2010 {\it Physiology} {\bf 25} 50--56

\bibitem{454} Sowerby S J, Broom M F and Petersen G B 2007 {\it Sensors and Actuators B} {\bf 123} 325--330

\bibitem{660} Willmott G R and Moore P W 2008 {\it Nanotechnology} {\bf 19} 475504
  
\bibitem{738} Willmott G R, Broom M F, Jansen M L, Young R M and Arnold W M Resizable elastomeric nanopores {\it Molecular- and Nano-Tubes} (Berlin: Springer) (in press)
  
\bibitem{771} Willmott G R and Young R 2009 {\it AIP Conf. Proc.} {\bf 1151} 153--156

\bibitem{870} Willmott G R and Bauerfeind L H 2009 Detection of polystyrene sphere translocations using resizable elastomeric nanopores {\it Industrial Research Limited} report 2385 (arXiv:1002.0611v1)


\bibitem{882} Willmott G R, Yu S S C and Vogel R Pressure dependence of particle transport through resizable nanopores (submitted)


\bibitem{753} Lee S, Zhang Y, White H S, Harrell C C and Martin C R 2004 {\it Anal. Chem.} {\bf 76} 6108--6115

\bibitem{759} Han A, Sch\"{u}rmann G, Mondin G, Bitterli R A, Hegelbach N G, de~Rooij N F and Staufer U 2006 {\it Appl. Phys. Lett.} {\bf 88} 093901

\bibitem{769} Sun L and Crooks R M 1999 {\it Langmuir} {\bf 15} 738--741

\bibitem{503} Gasparac R, Mitchell D T and Martin C R 2004 {\it Electrochimica Acta} {\bf 49} 847–850

\bibitem{892} Sparreboom W, van~den Berg A and Eijkel J C T 2009 {\it Nat. Nanotechnol.} {\bf 4} 713--720

\bibitem{517} Rice C L and Whitehead R 1965 {\it J. Phys. Chem.} {\bf 69} 4017--4025

\bibitem{564} Grossman P D and Colburn J C 1992 {\it Capillary Electrophoresis} (San Diego: Academic Press)

\bibitem{780} Schulz S F, Gisler T, Borkovec M and Sticher H 1994 {\it J. Colloid Interf. Sci.} {\bf 164} 88--98

\bibitem{522} Cervera J, Alcaraz A, Schiedt B, Neumann R and Ram\'{\i}rez P 2007 {\it J. Phys. Chem. C} {\bf 111} 12265--12273

\bibitem{535} Constantin D and Siwy Z S 2007 {\it Phys. Rev. E} {\bf 76} 041202

\bibitem{671} Ram\'{\i}rez P, Apel P Y, Cervera J and Maf\'{e} S 2008 {\it Nanotechnology} {\bf 19} 315707

\bibitem{891} Riddick T M 1968 {\it Control of Colloid Stability through Zeta Potential} (New York: Zeta-Meter Inc.)

\bibitem{888} Miller C R, Vogel R, Surawski P P T, Jack K, Corrie S R and Trau M 2005 {\it Langmuir} {\bf 21} 9733--9740

\bibitem{889} Vogel R, Surawski P P T, Littleton B N, Miller C R, Lawrie G A, Battersby B J and Trau M 2007 {\it J. Colloid Interf. Sci.} {\bf 310} 144--250

\bibitem{890} Corrie S R, Vogel R, Keen I, Jack K, Kozak D, Lawrie G A, Battersby B J, Fredericks P and Trau M 2008 {\it J. Mater. Chem.} {\bf 18} 523--529

\bibitem{648} Hepburn C 1982 {\it Polyurethane Elastomers} (Barking: Applied Science Publishers)

\end{thebibliography}

\end{document}